\documentclass{ws-ijmpd}
\usepackage[super,compress]{cite}
\usepackage{aas_macros}
\usepackage{url,hyperref}

\begin{document}

\markboth{Gabici et al.}
{The origin of Galactic cosmic rays}

\catchline{}{}{}{}{}
%
%%%%%%%%%%%%%%%%%%%%%%%%%%%%%%%%%%%%%%%%%%%%%%%%%%%%%%%%%%%%%%%%%%%%

\title{THE ORIGIN OF GALACTIC COSMIC RAYS: \\ CHALLENGES TO THE STANDARD PARADIGM}

\author{STEFANO GABICI}

\address{APC, AstroParticule et Cosmologie, Universit\'e Paris Diderot, CNRS/IN2P3, CEA/Irfu, Observatoire de Paris, Sorbonne Paris Cit\'e, \\ 10, rue Alice Domon et L\'eonie Duquet, 75205 Paris Cedex 13, France \\ gabici@apc.in2p3.fr}

\author{CARMELO EVOLI}

\address{Gran Sasso Science Institute, Viale Francesco Crispi 7, 67100 L'Aquila, Italy \\ INFN, Laboratori Nazionali del Gran Sasso (LNGS), 67100 Assergi, L'Aquila, Italy}

\author{DANIELE GAGGERO}

\address{Instituto de F\'isica Te\'orica UAM-CSIC,
Campus de Cantoblanco, E-28049 Madrid, Spain}

\author{PAOLO LIPARI}

\address{INFN, Sezione Roma ``Sapienza'', piazzale A.~Moro 2, 00185 Roma, Italy}

\author{PHILIPP MERTSCH}

\address{Institute for Theoretical Particle Physics and Cosmology (TTK), RWTH Aachen University, 52056 Aachen, Germany}

\author{ELENA ORLANDO}

\address{Kavli Institute for Particle Astrophysics and Cosmology and Hansen \\ Experimental Physics Laboratory, Stanford University, Stanford, CA 94305, USA}

\author{ANDREW STRONG}

\address{Max Planck Institut f\"ur extraterrestrische Physik, Postfach 1312, D-85741 Garching, Germany}

\author{ANDREA VITTINO}

\address{Institute for Theoretical Particle Physics and Cosmology (TTK), RWTH Aachen University, 52056 Aachen, Germany }

\maketitle

%\begin{history}
%\received{Day Month Year}
%\revised{Day Month Year}
%\end{history}

\begin{abstract}
A critical review of the standard paradigm for the origin of Galactic cosmic rays is presented. Recent measurements of local and far-away cosmic rays reveal unexpected behaviours, which challenge the commonly accepted scenario. 
These recent findings are discussed, together with long-standing open issues.
Despite the progress made thanks to ever-improving observational techniques and theoretical investigations, at present our understanding of the origin and of the behaviour of cosmic rays remains incomplete. 
We believe it is still unclear whether a modification of the standard paradigm, or rather a radical change of the paradigm itself is needed in order to interpret all the available data on cosmic rays within a self-consistent scenario.
\end{abstract}


\ccode{PACS numbers: 96.50.S, 96.50.Tf, 13.85.Tp, 95.85.Ry}

%\tableofcontents

\section{Introduction}
\label{sec:intro}

The problem of the origin of Cosmic Rays (CRs) is a central one in high energy astrophysics. 
While it is firmly established that the bulk of CRs originates within the Galaxy, the way in which these particles are accelerated at their sources, as well as the way in which they are confined in the magnetized and turbulent interstellar medium (ISM) are still a matter of debate.

An impressive amount of data of ever improving quality has been accumulated over a century of direct and indirect observations of CRs. According to the mainstream interpretation of these data, Galactic CR nuclei are believed to be accelerated at supernova remnant (SNR) shocks via first order Fermi mechanism and to be then somehow released into the ISM, where they reside for some time before escaping the Galaxy.

The goal of this paper is to provide a critical review of this standard paradigm for the origin of Galactic CRs. In the following, we describe in some detail the main aspects of the paradigm (Sec.~\ref{sec:orthodoxy}), and we review the most recent direct and indirect observations of CRs that are in tension with it (Sec.~\ref{sec:recentobservations}).
In Sec.~\ref{sec:problems} we list the major difficulties encountered by the paradigm in explaining observations, and we conclude in Sec.~\ref{sec:conclusions}.

For a more extended review on CRs the reader is referred to the excellent monographs \cite{1964ocr..book.....G,Ginzburg:1990sk,gaisser2016cosmic}.
The content of this review is based on a series of discussions that took place at the ''The High-Energy Universe: Gamma-Ray, Neutrino, and Cosmic-ray Astronomy'' MIAPP workshop in 2018. See \cite{2019arXiv190306714A} for a companion paper on Ultra-High Energy CRs.

\section{The orthodoxy}
\label{sec:orthodoxy}

With {\it orthodoxy} we refer here to the ideas and assumptions which are most commonly invoked to interpret the observations of CRs. 
These ideas and assumptions are briefly reviewed in the remainder of this Section, and constitute a standard framework for CR studies which is broadly (though not unanimously!) accepted. 
In the following, we identify what are in our view the three pillars of the orthodoxy (Sections \ref{sec:pillar1} to \ref{sec:pillar3}), and we discuss then two classical tests for CR origin based on very-high, and high-energy gamma-ray observations of SNRs, respectively (Sec.~\ref{sec:test} and \ref{PionBump}). 
We focus here onto CR nuclei only, and we postpone the discussion of CR electrons and antiparticles to Sec.~\ref{sec:recentobservations} and \ref{sec:problems}.

\subsection{The bulk of the energy of cosmic rays originates from supernova explosions in the Galactic disk}
\label{sec:pillar1}

The measured local energy density of CRs is $w_{CR}^0 \sim 1$ eV/cm$^3$, and is roughly equal to the energy densities of the other components of the ISM (magnetic field, soft photon backgrounds, thermal gas and turbulent motions), although this does not necessarily imply energy equipartition in a physical sense.
Before the recent observations performed by PAMELA and AMS02, the observed local spectrum of CR nuclei (largely dominated by protons) appeared to be consistent with a single power law in particle rigidity $R = p c / (Z \mathrm{e})$, $n_{CR}(R) \propto R^{-\nu}$ of slope $\nu \approx 2.7$ spanning the remarkably broad energy range from the multi-GeV domain to a few PeV. The steepness of the spectrum implies that most of the CR energy density is carried by $\gtrsim$ GeV protons and Helium nuclei.
The Larmor radius of these particles, in the typical magnetic field of few $\mu\text{G}$ found in the ISM, is of the order of $\sim 10^{12}$ cm, which is many orders of magnitude smaller than any typical Galactic length scale (e.g. coherence length of the ISM magnetic field, radius of the Galactic disk, etc.). This fact %and the energy equipartition of CRs with the other components of the ISM 
fits nicely with a scenario where CRs are accelerated within the Galaxy and effectively confined there. 

The confirmation of a Galactic origin of the bulk of CRs at least at GeV energies came from gamma-ray observations.
The bright diffuse gamma-ray emission detected in the GeV domain from the gaseous Galactic disk can be explained as the result of the decay of neutral pions produced in CR interactions with the gas, provided that the CR energy density and spectrum are {\it roughly} uniform throughout the entire Galactic disk. If CRs were {\it universal} (i.e. the CR intensity was the same everywhere in the Universe) rather than Galactic one would expect to detect an analog diffuse-gamma ray flux from external galaxies which would simply scale with the gaseous mass of the galaxy \cite{1976RvMP...48..161G}. 
The upper limit derived by EGRET on the gamma-ray flux from the Small Magellanic Cloud ruled out this possibility, and pointed towards a Galactic origin of the bulk of CRs \cite{1993PhRvL..70..127S}. Fermi-LAT observations of LMC, SMC and M31 also confirm that the CR intensity is different from the Galaxy \cite{2010A&A...523L...2A}.

Even though a small radial gradient of the CR intensity in the Galaxy is indeed observed (and expected), to a reasonably good approximation the total energy in form of CRs in the disk can be estimated as $W_{CR} \sim w_{CR}^0 \times V_d \approx 10^{55}$~erg, where $V_d$ is the volume of the disk.
To maintain a steady-state, the power that Galactic sources have to inject in the Galaxy in form of CRs is $P_{CR} \sim W_{CR}/\tau_{res}$, where $\tau_{res}$ is the residence time of CRs in the disk and where we implicitly assumed that CRs sources are located within the disk.

The way in which the residence time $\tau_{res}$ is estimated is connected to striking features in the chemical composition of CRs \cite{gaisser2016cosmic}. The light elements Li, Be, and B, and the sub-iron elements (Sc, Ti, V, Cr, and Mn) are overabundant by a many order of magnitude in CRs with respect to the solar system. 
This fact can be explained if such CR elements are {\it secondaries}, i.e., are produced in spallation reactions between primary CR nuclei and the ISM. 

In order to understand in a simple (and simplified) manner how $\tau_{res}$ is extracted from CR data, consider 
a species $s$ of CRs which is entirely secondary in nature (boron, for example), and that is mostly produced in spallation reactions involving a heavier CR species $p$ which, conversely, is mostly primary (like carbon).
The production rate of species $s$ can be written as $q_s(\epsilon) \sim n_p(\epsilon) n_{ISM} \sigma_s c$, where $n_{ISM}$ is the density of the ISM, $\sigma_s$ is the relevant spallation cross section, and $c$ is the speed of light. 
In the expressions above, $\epsilon$ is the particle energy per nucleon, which is a (almost) conserved quantity in spallation reactions, and therefore will be omitted in the following.
The equilibrium density of secondary CRs in the disk is then $n_s = q_s \tau_{res}$, which can be recast in terms of the {\it grammage} $\Lambda = \mu m_p n_{ISM} \tau_{res} c$ accumulated by CRs while residing in the Galactic disk, to eventually derive the important relation $n_s/n_p \sim \sigma_s \Lambda / \mu m_p$ for the secondary-to-primary ratio, which is an observable quantity.
In the expressions above $\mu$ is the mean atomic weight of the ISM ($\mu \sim 1.4$ for a hydrogen gas with 10\% of helium) and $m_p$ the proton mass. 
Therefore, from the knowledge of the spallation cross-sections and from the measurement of the CR secondary-to-primary ratios (for example the B/C ratio), it is possible to conclude that GeV CRs have to accumulate a grammage of $\Lambda \approx 10$~g/cm$^2$ while residing within the disk. 
For the typical density of the ISM $n_{ISM} \sim 1$~cm$^{-3}$ this translates into a residence time {\it in the Galactic disk} $\tau_{res}$ of the order of few Myr. 

From the estimate of the residence time in the disk it is possible to compute the CR power of the Galaxy as $P_{CR} \sim 10^{41}$~erg/s \cite{2010ApJ...722L..58S}.
It has to be stressed that the equivalence between grammage and \emph{Galactic} residence time is a very strong (and not necessarily fully justified) assumption which constitutes a cornerstone of orthodoxy. 
Radically different pictures could be, in principle, envisaged. For example, the grammage could be entirely or partially accumulated by CRs inside or in the vicinity of their accelerators, rather than during their residence in the ISM \cite{1973ICRC....1..500C}. Under these circumstances, the residence time in the disk would be decoupled from the grammage.

The suggestion that supernova explosions could provide the energy needed to explain CRs was first proposed by Baade \& Zwicky in 1934 \cite{1934PNAS...20..259B}, even though their reasoning is now outdated (at that time CRs were thought to be extragalactic). The connection with {\it Galactic} supernovae was made after noticing that the total injection rate of mechanical energy into the ISM would be the product of the supernova explosion rate in the galaxy (about 3 per century) with the typical mechanical energy released during one of such explosions (the canonical $10^{51}$ erg). This amounts to $\sim 10^{42}$ erg/s and significantly exceeds $P_{CR}$.
Thus, a very plausible 10\% efficiency in the conversion between supernovae kinetic energy and CR energy has to be invoked in order to make this hypothesis viable \cite{1950RvMP...22..119T,1964ocr..book.....G}.

Sources of energy other than stellar explosions have been proposed over the years to explain CRs, including explosive phenomena in the Galactic centre \cite{1964ocr..book.....G,1981SvA....25..547P} and stellar winds \cite{1980ApJ...237..236C}. However, these scenarios seemed to be more problematic, and the assumption that CRs acquire their energy from Galactic stellar explosions became almost universally accepted very soon.
Interestingly, as we will see in the following, both the Galactic centre and stellar wind scenarios for the origin of Galactic CRs have been recently revived based on gamma-ray observations.

\subsection{Cosmic rays are diffusively confined within an extended and magnetized Galactic halo}
\label{sec:pillar2}

A residence time in the Galactic disk $\tau_{res}$ of the order of few million years corresponds to a path traveled by CRs roughly equal to $\lambda_{res} = \tau_{res} c \sim 1$~Mpc. This length largely exceeds both the radius and the thickness of the disk. 
A possible way to reconcile these numbers is to assume that the motion of CRs is diffusive rather than rectilinear in the disk. 

In fact, another major constraint on CR propagation in the Galaxy can be obtained through the study of unstable secondaries such as the radioactive isotope $^{10}$Be.
This isotope is particularly suitable for these studies because its decay time, $\tau(^{10}{\rm Be}) \sim 1.4$ Myr, is of the same order of the residence time of CRs in the disk $\tau_{res}$.
$^{10}$Be is produced together with stable Be isotopes during CR spallation reactions in the disk, and its decay means that the observed ratio $^{10}$Be/Be is suppressed with respect to the production ratio, approximately by a factor of $\approx \tau(^{10}{\rm Be})/\tau_{esc}$, where $\tau_{esc}$ is the escape time of CRs from the Galaxy.
The measured value of the $^{10}$Be/Be ratio implies that $\tau_{esc} \sim 10-20$ Myr \cite{gaisser2016cosmic}, which significantly exceeds the residence time of CRs in the disk $\tau_{res}$!

This apparent discrepancy is solved by assuming that CRs are diffusively confined within a magnetized Galactic halo, whose volume is much larger than that of the disk.
In this scenario, the effectiveness of spallation reactions is strongly reduced while CRs reside in the halo, due to the very low gas density there, and the grammage is entirely accumulated by CRs during repeated crossings of the thin gaseous Galactic disk.

An estimate of the typical value of the diffusion coefficient of CRs $D_0$ in the halo can now be obtained by considering the Galaxy (disk plus halo) as a box characterized by an average gas density $\bar{\varrho} = \mu m_p n_{ISM} (h/H)$, where $h \approx 100$ pc and $H$ (a priori unknown) are the thicknesses of the disk and halo, respectively. In such a box, the grammage accumulated over a time $\tau_{esc} \sim H^2/D_0$ is $\Lambda = \bar{\varrho} \tau_{esc} c$ which can be inverted to get:
\begin{equation}
D_0 \sim 3 \times 10^{28} \left( \frac{H}{5~{\rm kpc}} \right) \left( \frac{\Lambda}{10~{\rm g/cm^2}} \right)^{-1} {\rm cm^2/s} ~.
\end{equation}
The estimate of $D_0$ depends on the poorly constrained thickness of the halo, and has to be considered as a typical value representative of the entire volume of containment.

Such an estimate of the CR diffusion coefficient refers to the bulk of CRs, which have a particle energy of $\gtrsim 1$ GeV, but the same procedure can be repeated for particles of different energies, leading to an energy dependent estimate of both the escape time from the Galaxy $\tau_{esc}$ and the residence time in the disk $\tau_{res}$. 
Within this context, it is more appropriate to express the CR diffusion coefficient as a function of the particle rigidity $R$ (particles of the same rigidity follow the same trajectories through an arbitrary magnetic field) 
as: $D = D_0 \beta (R / R_0)^{\delta}$, 
where $\beta$ is the particle velocity in units of the speed of light. Various analyses of CR and gamma-ray data gave as best fit parameters $\delta \sim 0.3-0.6$ and $D_0 \sim 10^{28} - 10^{29}$ cm$^2$/s for $R_0$ equal to few GV \cite{2011ApJ...729..106T,2012ApJ...750....3A,2008JCAP...10..018E,2007ARNPS..57..285S}.

An important prediction of any diffusive model for CRs is the presence of a small dipole anisotropy in the arrival direction of CRs at a level of the order of $a = (3~D_0/c) |\nabla n_{CR}|/n_{CR}$ where $n_{CR}$ is the CR space density \cite{1964ocr..book.....G}. Taking $L$ to be the typical scale of the CR gradient over large Galactic scales one gets $a \approx 10^{-4} (D_0/10^{28}~{\rm cm^2/s}) (L/3~{\rm kpc})^{-1}$. 
However, the role of nearby sources of CRs has to be taken into account carefully, since it might dominate the anisotropy. Moreover, the very weak dependence of the observed anisotropy on particle energy seems to point towards an equally weak dependence on energy of the CR diffusion coefficient ($\alpha \approx 0.3$) \cite{2006AdSpR..37.1909P} (See Sec.~\ref{sec:anisotropies} for more details).

\subsection{Cosmic rays are accelerated out of the (dusty) interstellar medium through diffusive shock acceleration in supernova remnants}
\label{sec:pillar3}

In order to keep the CR intensity in the Galaxy at the observed level, the injection rate of CRs of a given energy $E$ in the ISM $q_{CR}(E) \propto E^{-\Gamma}$ should balance the rate at which CRs leave the Galaxy. This can be expressed as $q_{CR}(E) = n_{CR}(E)/\tau_{esc}(E) \propto n_{CR}(E) D(E) \propto E^{-\nu+\alpha}$. Given that $\nu \sim 2.7$ and $\alpha \sim 0.3-0.6$ one gets $q_{CR}(E) \propto E^{-2.1...2.4}$. 
Thus, CR sources have to inject particles in the ISM with a power law spectrum somewhat steeper than $E^{-2}$. Since such a spectral index is quite close to the test-particle prediction of diffusive shock acceleration theory \cite{1977SPhD...22..327K,1977ICRC...11..132A,1978MNRAS.182..147B,1978ApJ...221L..29B}, the idea that SNR shocks are the acceleration sites of Galactic CRs became very broadly accepted.

In fact, diffusive shock acceleration theory predicts, in the test particle limit, CR spectra at strong shocks which are power laws with slope {\it identical} to $E^{-2}$, and a number of theoretical arguments seem to indicate that this should be also the shape of the spectrum of the CRs which escape SNRs and are eventually injected in the ISM \cite{2005A&A...429..755P,2013MNRAS.431..415B}.
These predictions of diffusive shock acceleration theory are in tension with the requirements derived from the observations of local CRs described above and from the observations in the gamma-ray domain of SNRs \cite{2013APh....43...71A,2011JCAP...05..026C,2018MNRAS.475.5237G}, both pointing to spectra of CRs steeper than $E^{-2}$. This discrepancy might be reconciled by relaxing the test-particle assumption and considering the effects of CRs on the shock structure. 
One of the most prominent effects of the CR feedback on shocks is a dramatic amplification of the turbulent magnetic field \cite{2004MNRAS.353..550B}, which could in turn impact onto the acceleration process.
For example, if the magnetic field is significantly amplified, the Alfv\'en speed of the plasma increases accordingly. If Alfv\'en waves are the main scattering agents of CRs at shocks, their enhanced motion (called ``Alfv\'en drift'') may lead to the formation of particle spectra significantly steeper than $E^{-2}$ \cite{2008AIPC.1085..336Z,2009MNRAS.395..895C,2018MNRAS.474L..42R}. Also the effect of the presence of neutral particles in the gas swept up by the SNR shock may lead to a steepening of the spectrum of accelerated particles \cite{2012ApJ...755..121B}.
However, these theoretical predictions should be taken with some caution, given that a fully self-consistent picture of the acceleration and escape of particles from SNR shocks is still missing \cite{2011MmSAI..82..760G}.

Finally, any satisfactory theory of CR origin must explain the observed chemical composition of these particles. 

Overall, the CR chemical composition follows quite closely the solar one, with the above mentioned conspicuous exceptions of light and sub-iron elements, which are many orders of magnitude overabundant in CRs and result from nuclear spallation reactions of CRs with interstellar matter. Less pronounced, but still significant differences include the overabundance of CR nuclei with $Z > 2$ with respect to H and He, and of refractory elements over volatiles \cite{1997ApJ...487..182M}. Refractory elements are mainly found condensed in dust grains, while volatiles are mainly in the gas phase of the ISM.
While the former difference remains unexplained, the latter seems to be related to atomic properties, rather than nuclear processes \cite{2007SSRv..130..415W}. In particular, data can be explained satisfactorily by an acceleration mechanism whose efficiency increases with the particle mass-to-charge ratio, resulting in a preferential acceleration of charged dust grains over atomic nuclei.
In such a scenario the overabundance of refractory elements would result from the sputtering of accelerated dust grains at shocks (for details see \cite{1997ApJ...487..197E,2017PhRvL.119q1101C}). This demonstrates that dust plays a crucial role in the determination of the chemical composition of Galactic CRs.

\subsection{Very-high-energy gamma rays from supernova remnants: a classic test for cosmic ray origin}
\label{sec:test}

A classic test of the supernova paradigm for the origin of CRs is based on gamma-ray observations of SNRs \cite{1994A&A...287..959D}. SNRs can explain CRs if on average $\approx 10$ \% of each parent supernova explosion energy ($\sim 10^{51}$ erg) is converted into accelerated particles with a power law spectrum in energy with slope slightly steeper than 2. If these conditions are satisfied one can predict in an almost model-independent way the typical gamma-ray luminosity of SNRs due to proton-proton interactions of accelerated CRs with the ambient medium compressed at the SNR shock. Such a prediction will depend only on the ambient density in the Galactic disk, whose average value is $n_{\rm ISM} \sim 0.1 ... 1$ cm$^{-3}$. 

Several SNRs have been detected in gamma rays both in the GeV \cite{2016ApJS..224....8A} and TeV \cite{2018A&A...612A...3H} energy domain, with gamma-ray fluxes comparable to the above mentioned predictions. However, the detection of gamma rays from SNRs does not necessarily mean that CR protons are accelerated in these objects, because inverse Compton scattering of accelerated {\it electrons} off soft ambient photons might also explain observations. A prototypical example is the SNR RX~J1713.7-3946 \cite{2011ApJ...734...28A,2018A&A...612A...6H}, whose gamma-ray emission could be explained either in a leptonic \cite{2012ApJ...751...65F} or hadronic \cite{2014MNRAS.445L..70G} scenario.

The predictions of the SNR paradigm have been tested also against the results obtained from the survey of the Galactic plane performed by H.E.S.S., which covers the interval of Galactic coordinates $350^{\circ} < l < 65^{\circ}$ and $|b| < 3^{\circ}$ down to a TeV gamma-ray flux level of $\lesssim 1.5$\% Crab units \cite{2018A&A...612A...1H}. The number of firmly identified SNRs in the survey is 8, which is consistent with the expectations of the SNR paradigm \cite{2013MNRAS.434.2748C}. Even though the agreement between predictions and observations is encouraging, it should be kept in mind that such agreement constitutes a necessary but not sufficient conditions for the validity of the SNR paradigm. First of all, the exact number of SNRs detected in the survey is quite uncertain. In addition to the firm associations, 8 more objects have been associated to {\it composite} SNRs, which exhibit both a shell-like and a pulsar wind nebula-like emission. Moreover, more than half of the sources are not firmly identified or not identified at all (47 over 78). Secondly, the relative contribution to the gamma-ray emission from pion decay and inverse Compton scattering remains uncertain, and a significant contribution from the latter cannot be ruled out. 
Future observations of SNRs with the \v{C}erenkov Telescope Array will certainly reduce such uncertainties, and might finally prove or falsify the SNR paradigm for the origin of CRs \cite{2017MNRAS.471..201C}.

The advent of gamma-ray facilities of superior sensitivity such as the \v{C}herenkov Telescope Array will also pave the way for another promising means to reveal the acceleration of hadrons at SNRs. This can be done by searching for the gamma-ray emission produced {\it in the vicinity} of SNRs due to proton-proton interactions between CRs escaping the SNR shock (runaway CRs) and the surrounding medium. Such hadronic emission is amplified if a massive molecular cloud is located in the proximity of the SNR. To date, only at most a couple of clear cases of molecular clouds illuminated by runaway CRs are known, and increasing the statistics of detections is of paramount importance since this might provide a proof for the acceleration of CR hadrons at SNRs (see \cite{2009MNRAS.396.1629G} or \cite{2015icrc.confE..29G} for a review).

\subsection{High-energy gamma rays from SNRs: the pion-bump}
\label{PionBump}

As seen above, many SNRs have been studied in gamma rays, with excellent spectral information from both satellite and \v{C}erenkov instruments. Frequently the phrase `pion bump' or 'pion peak' is used, referring to the maximum in the $\pi^0$-decay spectrum from hadronic interactions, and cited as `proof' of hadronic CRs in SNRs. 

However this is very misleading: the peak is at one-half of the $\pi^0$ mass, i.e. 67.5 MeV, and the (differential) gamma-ray energy spectrum is in fact symmetrical on a log-energy scale independent of the proton (and Helium) spectra which determine the $\pi^0$ spectrum. 
This is a simple consequences of the isotropic decay in the $\pi^0$ rest frame and the Lorentz transformation. A detailed analysis can be found in the book by Floyd Stecker\footnote{Cosmic Gamma Rays, available at\\ https://ntrs.nasa.gov/archive/nasa/casi.ntrs.nasa.gov/19710015288.pdf, see Chapters 1-6 and 5-2}.
The peak energy is however below the limit reached by Fermi-LAT in current analyses, which is about 100 MeV; to really see the peak would need a spectral measurement extending below the maximum, say to 40 MeV at least. The AGILE instrument has spectra extending to slightly lower energies but the same argument applies.

Why then is the pion bump often invoked? It seems that it is because the spectrum is normally presented with a factor $E^2$ to reflect the energy distribution and improve the readability of the data (which is quite in order and standard practice). Then the spectra do indeed often show a peak around 1 GeV, and this seems often thought to be (hearing people's comments) the shifted pion bump to higher energies due to the $E^2$ factor. 
However if the peak is not in the raw differential flux spectrum, multiplying by this factor cannot make it appear! The observed peak is in fact usually due to the fact that the gamma-ray spectrum reflects a proton spectrum with a break in the several GeV range.

To illustrate this explicitly, in \cite{STRONG2018165} two sample SNR spectra were taken from the literature (W44 \cite{2014A&A...565A..74C,2016A&A...595A..58C} and W49B \cite{2016arXiv160900600H}) and replotted with the $E^2$ factor removed. Then there is no pion bump visible since it lies below the minimum energy measured.

Of course the proper procedure is to make explicit models of the hadronic and leptonic emission, and there are many examples where this is done, but the interpretation is necessarily model-dependent. 
An example where the combination of gamma-ray and synchrotron data favours a mainly leptonic model is RCW 86 \cite{2016ApJ...819...98A}. The `smoking gun' pion bump will have to await extension of Fermi-LAT analysis to lower energies, which is indeed foreseen in the new `Pass 8' event analysis (although this is difficult because of the broad angular response at low energies), and future experiments (e.g. COSI, eAstrogam) which extend into the low MeV range.

\section{Recent observations confront orthodoxy}
\label{sec:recentobservations}

In this Section we review a number of recent direct and indirect observations of Galactic CRs.
We put particular emphasis onto those observations that constitute a challenge to the standard scenario described above.

\subsection{Spectral anomalies in recent observations of local cosmic rays}
\label{sec:spectral}

\subsubsection{Spectral anomalies for cosmic ray nuclei}
\label{sec:nuclei}

The last few years have seen improvements in our knowledge of CR energy spectra.
In particular, the orthodox picture of a universal injection spectrum of Galactic sources (as far as primary species are concerned) and a power law scaling of the CR diffusion coefficient with energy, both without any features up to the CR knee, has been undermined by more precise measurements of individual CR spectra by a number of CR experiments: ATIC \cite{Panov2009}, AMS~\cite{PhysRevLett.114.171103,PhysRevLett.115.211101}, BESS~\cite{2016ApJ...822...65A}, CREAM~\cite{Yoon:2011aa} and PAMELA~\cite{2011Sci...332...69A}.

In particular, PAMELA first, and AMS02 later on with increasing precision, have robustly highlighted a new situation with respect to what was routinely assumed before then. 
The two key observational results that these experiments have pointed out are: 
\begin{itemize}
\item The propagated proton spectrum is distinctly softer ($\Delta \gamma \sim -0.1$) than that of helium at all energies.
\item Both spectra show a spectral hardening (with change in spectral index of $\sim 0.13$) around $\sim$300 GV.
\end{itemize}

Carbon and oxygen are the most abundant nuclei heavier than helium in CR fluxes. 
Recently AMS02 reported precise measurements of their fluxes up to 3 TV. The rigidity dependence of the helium, carbon, and oxygen spectral indices are very similar. In particular, all spectral indices harden at around the same rigidity and with the same change of slope as for H and He. Additionally, the hardening allows one to reconcile earlier low-energy measurements with the high-energy trends reported by CREAM. 

The spectral indices of carbon, oxygen and helium are identical within the measurement errors above $\sim$60 GV, meanwhile the ratios He/O and C/O exhibit a soft decrease with rigidity below that value. 
These trends confirm expectations based on the standard picture in which CRs are injected with the same spectrum at the source and are affected by transport effects in the ISM in the same way. The only (tiny) difference among the primary species is in their fragmentation rate (roughly proportional to $A^{0.7}$) and that Carbon has a not-negligible secondary contribution from Oxygen fragmentation~\cite{Evoli:2019}.

The spectral difference between protons and helium is still not understood and several explanations have been put forward.

In terms of transport, the only difference between the two species is that helium fragments in the ISM.
In fact, at energies where the helium fragmentation is faster than galactic escape the helium spectrum is expected to be slightly harder than proton one. However, $^4$He spallation is dominated by fragmentation into $^3$He which does not harden the total He spectrum.
It can been shown that, when the escape time is set to reproduce the B/C, this effect should disappear at $\sim$50~GeV and therefore is not sufficient to explain the different slopes observed by AMS-02 up to $\sim 2$~TV~\cite{Evoli:2019}. 
Therefore, transport properties of CRs seem unable to explain the difference between H and He spectra, unless quite radical departures from the standard scenario are invoked, such as a decoupling of the escape timescale from the grammage needed to reproduce B/C. 

At the moment it is then reasonable to assume that the solution to this puzzle is a matter of acceleration or escape from the sources. 
Specific modifications of the diffusive shock acceleration mechanism have been proposed. 
A spectral difference between protons and helium is expected if they are injected and accelerated at different stages of the source evolution.
These scenarios may involve a shock whose velocity decreases with time. Thus, helium must be mostly accelerated early on (when the Mach number $M$ is much larger than 1, and the spectrum of accelerated particles is harder) while protons should be injected predominantly when $M$ is lower~\cite{2015JPhG...42g5201E}. 
That could be the case for non-uniform helium distribution in the medium surrounding the accelerators~
\cite{2011ApJ...729L..13O}. Even though these elements are omnipresent in the ISM with almost constant abundance ratio, such a particular configuration might be achieved inside superbubbles. Another class of models assume that He ions are preferentially injected with respect to protons when the shock is stronger given the larger He gyroradius downstream the shock~
\cite{2012PhRvL.108h1104M}. This possibility is still under debate given our poor knowledge of the injection efficiency as a function of environmental parameters. To this end, it has been shown by means of numerical simulations in~\cite{2018arXiv180300428H} that the efficiency of injection increases with A/Z and peaks around $M \sim 5$ such that, when convoluted with the shock time dependence, it allows to reproduce the AMS-02 data.

From all these mechanisms, the p/He ratio is expected to decrease steadily up to the maximum rigidity attainable by the accelerators. However,~\cite{2015ApJ...815L...1T} has noticed that the existing data at higher energies show no evidence of spectral differences between protons and He, although the large discrepancy among the data from different experiments does not allow to reach firm conclusions.
Alternative explanations able to reproduce the high-energy flattening for the p/He ratio, are based on distinct populations of sources, hydrogen rich ones producing rather softer spectra (presumably a nearby SNR characterized by a soft acceleration spectrum) and helium rich ones producing systematically harder ones~\cite{2015ApJ...815L...1T}. 

Let us consider now the spectral break (hardening) observed for primary nuclei at a rigidity of $\sim$300 GV. This feature could be explained either in terms of transport properties of CRs in the Galaxy (e.g. a feature in the CR diffusion coefficient)
~\cite{2013JCAP...07..001A,2012PhRvD..85l3511E,2012ApJ...752L..13T}, or in terms of a break in the overall spectrum of CRs injected in the ISM~\cite{2013ApJ...763...47P,2006A&A...458....1Z,2011PhRvD..84d3002Y,2018MNRAS.474L..42R}. 

The question about the origin of the break can find an answer thanks to precise measurements of secondary spallation nuclei. In fact, if it were due to a pure propagation effect, the break in the secondary spectra would be twice as pronounced than that in the primary spectra, while it should be the same if the break is present in the injection spectrum. 
In~\cite{2017PhRvL.119x1101G}, using boron to carbon ratio data recently released by AMS02 experiment up to $\sim$TeV energies, the authors found evidence in favor of a diffusive propagation origin for the broken power-law spectra found in primary species. More recently, the AMS02 measurements of the absolute spectra of secondary species, Li, Be and B, confirm that the spectral slopes above the break are compatible with this hypothesis~\cite{2018PhRvL.120b1101A}. 

\begin{figure}[t]
\centering
\includegraphics[width=\textwidth]{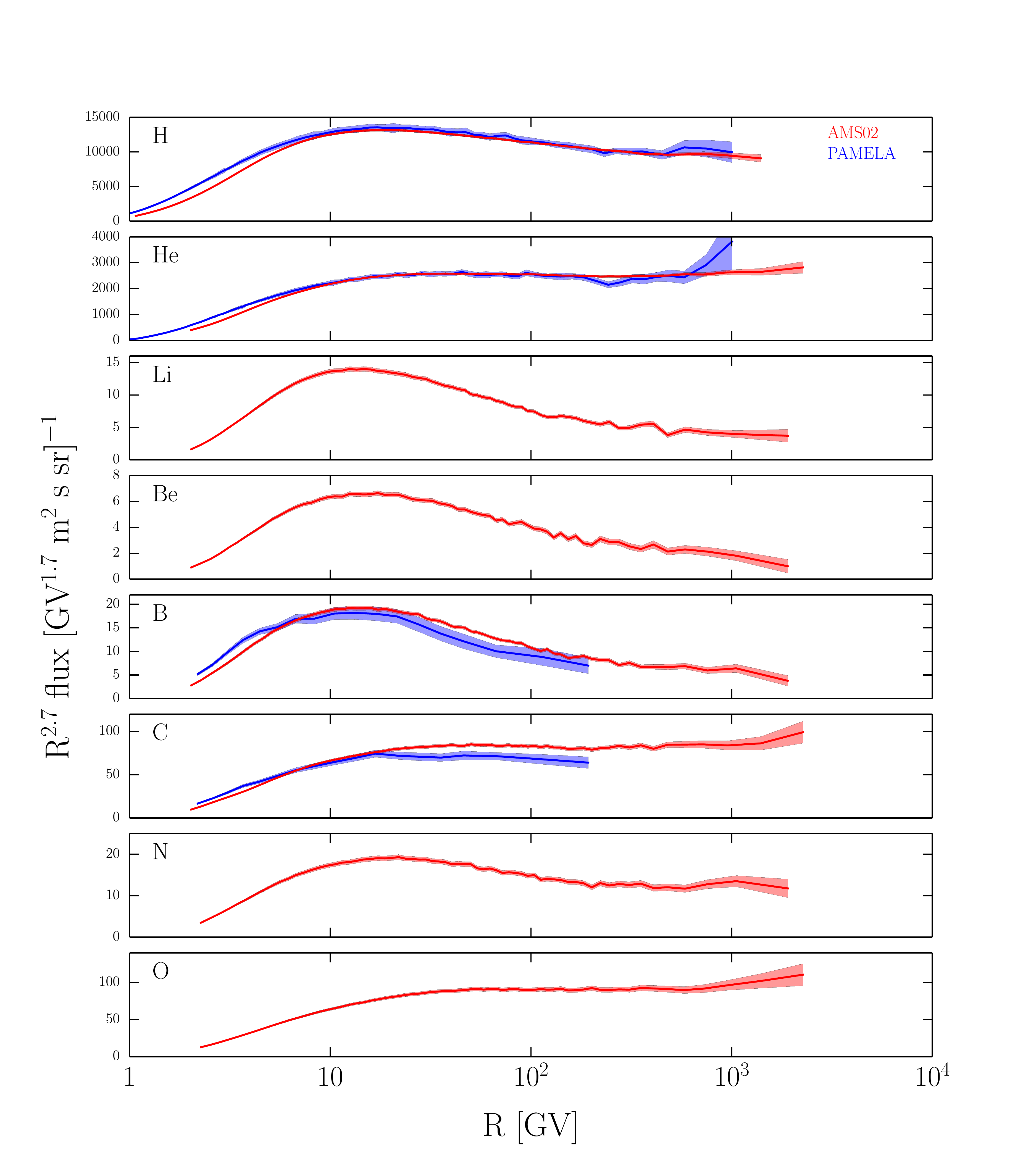}
\caption{The individual CR flux for nuclear species up to Oxygen as measured by PAMELA and AMS02. Shadow regions correspond to 1 sigma total errors (systematic and statistical added in quadrature). 
}
\end{figure}

The origin of the feature in the diffusion coefficient invoked to reproduce the data could have different origins. 
In~\cite{2012ApJ...752L..13T} it was assumed that the diffusion coefficient is a non-factorisable function of rigidity and space and, in particular, the rigidity dependence of diffusion in the region close to the Galactic disk is different than in the outer diffusive halo. 
These models can find a physical justification in the fact that in the inner zone SNR driven turbulence should dominate, while the turbulence in the halo (which is more relevant for low-energy CRs) is of different origin. 

An alternative model is the one introduced in~\cite{2012PhRvL.109f1101B}, where the turbulence scattering CRs is described by two components: the external turbulence (as traditionally argued to be injected by SNRs) with a Kolmogorov spectrum, and the waves generated by CRs themselves via the streaming instability~\cite{1969ApJ...156..445K}. 
In particular, CRs above the break diffuse on external turbulence, while self-generated turbulence dominates at lower rigidities.
A simple estimate shows that the non-linear damping rate of Alfv\'en waves equals the CR streaming instability growth rate at a rigidity of 200-300 GV, for fiducial values of Galactic parameters, tantalizingly close to the rigidity of the break. Remarkably, this fact was noticed in \cite{2004ApJ...604..671F} well before PAMELA and AMS02 data were available.
A more complete treatment, including wave advection from the Galactic disc where the sources of CRs and turbulence are assumed to be located, showed that the halo naturally arises from these phenomena, with a size of a few kiloparsecs, compatible with the value that typically best fits observations in parametric approaches to CR diffusion~\cite{2018PhRvL.121b1102E}. 

These models derive from an improved microphysics description of the CR propagation problem, however this is not yet well known and different aspects (e.g., the presence of neutral region where the waves are rapidly damped, the anisotropy of the turbulent cascade, etc.) must be clarified.

\subsubsection{The peak in the B/C ratio: how important is diffusive reacceleration of cosmic rays?}
\label{Reacc}

The CR secondary-to-primary ratio, in particular B/C, shows a peak in energy at a few GeV. 
In older times this was modelled with an ad-hoc break in the grammage or diffusion coefficient, but a physical origin has been proposed 
as the diffusion in momentum which is expected at some level if the scattering entities for spatial diffusion are also moving.
This is referred to as diffusive reacceleration (DR).
For a detailed account see \cite{2017A&A...597A.117D}.
See also \cite{2014MNRAS.442.3010T} where a simple derivation of the approximate relation between spatial and momentum diffusion coefficients at momentum $p$:
 $ D_{xx}D_{pp} \approx {p^2 V_a^2/9}$ (where $V_a$ is the Alfv\'en speed) as well as the full formulae are given.
 
While DR can give a good fit to B/C and other secondary-to-primary ratios, and allows a more Kolmogorov-like exponent (1/3) in the $D_{xx}(p)$ law, and hence helps to avoid too large anisotropy when extrapolated to higher energies,
the question arises whether it really occurs at the level required for this to work.

In \cite{2017A&A...597A.117D} both analytical and numerical (GALPROP) methods were used to estimate the energy injection from DR for typical values of the parameter $V_a\approx$ 30 km s$^{-1}$ required by B/C, 
and found that 30-50\% of the total energy of CR then actually comes from the ISM via DR, and not from the usual sources such as SNR!
This refers to energies around a few GeV where most of the CR energy is concentrated.

So while not excluded, our conclusion is that DR should not be invoked just because it gives a good fit to CR data, but should be considered critically along with other possible origins for the shape of B/C. 

These include convection via a Galactic wind, which produces an energy-independent term in the propagation equation and hence dominates at low energies where diffusion becomes small under the usual momentum-dependent law $D_{xx}(p)= \beta p^\delta$ where $\delta=0.3-0.5$.
Note that in any case the velocity term in the secondary-production rate means that B/C decreases below 1 GeV as $\beta =$ v/c $< 1$,
(although this may be partly cancelled by the velocity term in $D_{xx}(p)$) and that the observed B/C has the rather uncertain effect of solar modulation as well. 

Another scenario that allows to reproduce B/C data features as a key ingredient an effective damping of the diffusion coefficient at energies around $\sim$ 1 GeV/n. In that picture, secondary/primary ratios can be reproduced with a smaller contribution of diffusive reacceleration and with a Kraichanan-like behaviour of the diffusion coefficient. The damping of the diffusion coefficient at the B/C peak has been introduced on phenomenological basis to reproduce the data in~\cite{2010A&A...516A..67M} and~\cite{2010APh....34..274D}. From a theoretical point of view, such behaviour is expected in the context of models that capture the resonant interaction between CRs and MHD waves, which results in a significant
dissipation of such waves~\cite{2006ApJ...642..902P,2014ApJ...782...36E}. As pointed out in \cite{yuan2017ams}, different effective parametrizations of the CR transport problem, including either strong reacceleration, advection, or an alteration of the diffusion coefficient at low energy, can provide reasonably good overall fits of all the hadronic CR fluxes measured by AMS-02.

While DR should occur at some level on physical grounds, it probably does not suffice to explain B/C; in any case experimental checks using other CR indicators would be valuable. Up to now we were unable to think of any critical tests however. Meanwhile other mechanisms like convection should be considered at least on equal terms with DR.

\subsubsection{A prominent break in the cosmic ray electron spectrum}
\label{sec:electrons}

The lepton component ($e^+$ + $e^-$) accounts for about 1\% of CRs. 
Secondary electrons and positrons are produced in interactions of CR nuclei with the interstellar gas, through the $\pi^\pm \rightarrow \mu^\pm + \mathellipsis \rightarrow e^\pm + \mathellipsis$ decay chains. 
Since approximately equal amounts of $e^-$ and $e^+$ are expected from such processes, the observed overabundance of $e^-$ over $e^+$ in CRs indicates that most of electrons have a primary origin. 
Moreover, observations of X-ray and TeV gamma-ray emission from SNRs provide clear evidence for the acceleration of electrons in SNR shocks up to energies of about $\sim 100$~TeV~\cite{2013APh....43...71A}.

The inclusive lepton ($e^+$ + $e^-$) spectrum was measured by \textit{Fermi}-LAT to follow a power-law with spectral index $-3.08 \pm 0.05$ in the energy range between 7 GeV and 2 TeV~\cite{2009PhRvL.102r1101A}.
More recently, with almost seven years of Pass 8 data, the measurements were tentatively fitted by a broken power-law with an energy break at $E_b = (53 \pm 8)$~GeV and spectral indices below and above the break of $\gamma = -3.21 \pm 0.02$ and $\gamma = -3.07 \pm 0.02$. 
The break is however not statistically significant when the systematic uncertainty on the energy measurement is taken into account.
As far as CR electrons are concerned, PAMELA was the first experiment to identify electrons above 50 GeV. 
The reported measurements of the $e^-$ spectrum between 30 and 625 GeV were well described by a single power-law with spectral index $-3.18 \pm 0.05$ and no significant spectral features within the errors.

More recently, AMS02 reported the most precise measurement to date of the fluxes of CR $e^-$, and $e^+$ + $e^-$ in~\cite{PhysRevLett.113.121102,PhysRevLett.113.221102}.
Consistent with previous measurements, no structures were observed in the $e^+$ + $e^-$ spectrum that can be described by a single power-law above 30 GeV with spectral index $\gamma = -3.170 \pm 0.008$(stat+syst)$\pm 0.008$ (energy scale).
On the contrary, the electron spectrum cannot be fitted with a single power-law over the energy range that is not affected by solar modulation ($>$10 GeV). 
In fact, it turned out that the spectral index $\gamma_{e^-}$ has an energy dependence and that the spectrum hardens at $\sim 30$~GeV. 

The observed electron spectrum is thus much steeper than that of protons. Unlike the hadronic CR component, electrons suffer significant energy losses when propagating in the Galaxy, but even accounting for this it is necessary to consider a different injection spectrum for electrons and protons~\cite{2011A&A...534A..54S,2013PhRvL.111b1102G,2018MNRAS.475.2724O} in order to reproduce the local electron spectrum. This might indicate either that CR protons and electrons are released by the same sources with a different spectrum, or that they are accelerated in different sources.

H.E.S.S. was the first experiment to extend the measurement of the lepton spectrum beyond the range accessible to direct measurements~\cite{2008PhRvL.101z1104A}. 
In~\cite{2009A&A...508..561A}, they reported evidence for a steepening in the energy spectrum at about 1 TeV. The most recently presented data show that the steepening is very sharp with the spectral index changing from $\gamma = -3.04$ to $\gamma = -3.78$~\cite{KerszbergICRC2017}. Moreover, the spectrum continues without further attenuation up to a particle energy of about 20 TeV. Observations by VERITAS confirmed the presence of the break \cite{2018PhRvD..98f2004A}. A very important implication of these observations is that, since the energy loss time of 20 TeV electrons in the Galaxy is extremely short, such particles must have been accelerated very recently. If this constraint is used together with the benchmark value of the CR diffusion coefficient (i.e. derived from B/C), one can conclude that 20 TeV electrons must have been produced in our very local neighbourhood ($\approx 100$ pc) \cite{1995A&A...294L..41A,2018arXiv181104123L,2018arXiv181107551R}.

Following that, CALET and DAMPE experiments performed this measurement up to a particle energy of 5 TeV with high level of precision providing further evidence for a break in the spectrum at around 1 TeV above which the spectral index rapidly changes from $\sim -3.2$ to $\sim -4.1$~\cite{2017PhRvL.119r1101A,2017Natur.552...63D}.

The sharp break seen by H.E.S.S., VERITAS, CALET, and DAMPE is one of the most pronounced feature observed in the spectrum of all CRs.
Regarding an interpretation, it was shown recently that the break in the all-electron spectrum can be interpreted in a stochastic model of sources~\cite{Mertsch:2018bqd}, or produced by a single, nearby and fading source of electrons \cite{2018arXiv181107551R}.

\subsubsection{The anti-matter component: unexpected behaviour of positrons and antiprotons}
\label{sec:anti}

Over the last decade, many experimental efforts have been aimed at measuring with high precision CR fluxes of anti-matter, including positrons and antiprotons.
Apart from important information about CR propagation, these particles provide a test for primary cosmological antimatter and for non-standard production.

With this respect, the result that drove most excitation in the community was the measurements of the CR positron fraction, $e^+ / (e^+ + e^-)$, by PAMELA~\cite{2009Natur.458..607A} (in fact, strengthening previous claim by HEAT~\cite{1997ApJ...482L.191B} and AMS-01~\cite{2007PhLB..646..145A}) and confirmed with impressive accuracy by AMS-02 from 0.5 to 500 GeV~\cite{PhysRevLett.110.141102}.

Unlike standard predictions for secondary production in ISM, the positron fraction exhibits a pronounced rise beyond $\sim 8$ GeV.
In parallel with measurements of the electron spectrum, such an increase has been interpreted as the signal for the presence of a primary source of positrons.
Other authors
\cite{Cowsik:2010zz,Cowsik:2013woa,Cowsik:2016wso,Katz:2009yd,Blum:2013zsa,Ahlen:2014ica,Lipari:2016vqk,Lipari:2018usj}
have however discussed scenarios where the positron flux is generated by the standard secondary production mechanism
(see also below the discussion in Sec.~3.1.4.)

Energy losses by IC scattering on CMB photons and synchrotron emission in the local magnetic field place an upper limit on the age ($\sim 10^5$ yr) and distance ($<1$~kpc) of the astrophysical sources of $\sim 100$~GeV positrons.
The observed excess of positrons was promptly interpreted as a long-waited signature of the presence of dark matter particles in the Milky Way halo~\cite{2008PhRvD..78j3520B}. 
However astrophysical sources as nearby and young pulsars~\cite{1995A&A...294L..41A,2009JCAP...01..025H,2009APh....32..140G} or nearby supernova remnants, which may be able to produce and accelerate secondary positrons before they escape into the ISM~\cite{2009PhRvL.103e1104B,2009PhRvL.103h1104M,Mertsch:2014poa}, have been shown to be able to maintain the observed population of these particles.

According to the latest AMS-02 data, the positron fraction likely drops above a particle energy of $\sim$ 400-500 GeV. This can be explained only if the local source(s) of TeV electrons do not produce positrons in equal amount, disfavoring scenarios involving pulsars/pulsar winds as the main sources of high energy leptons \cite{2018arXiv181107551R}.

Antiproton observations could be used to investigate the origin of primary positrons.
If the rise in the positron fraction is due to dark matter, antiproton data provide important constraints on annihilation or decay model~
\cite{2009PhRvL.102g1301D,2009NuPhB.813....1C,2012PhRvD..85l3511E}. 
At the same time, the SNR hypothesis leads as an unavoidable consequence to a secondary component with a hard energy spectrum, predicting a rise in the $\bar{p}/p$ ratio but also in the B/C, and the two cannot be accommodate consistently.

The first accurate measurements of the antiproton spectrum have been performed by CAPRICE98~\cite{2001ApJ...561..787B} and most notably by the BESS and BESS-Polar (1993-2008)~\cite{1997ApJ...474..479M,2000PhRvL..84.1078O,2008PhLB..670..103A} detectors.
More precise data have been collected by PAMELA~\cite{2010PhRvL.105l1101A,2013JETPL..96..621A} covering the range from 70 MeV to 200 GeV, and AMS-02~\cite{2016PhRvL.117i1103A} extending this range to 300 GeV. 

A fundamental characteristic of the antiproton spectrum is the presence of the peak around 2 GeV which is due to the kinematic threshold at $7 m_p \sim 6.5$~GeV of the main secondary production reaction: $p + p_{\rm ISM} \to 3p + \bar{p}$. 
The $\bar{p}$ spectrum falls sharply at low energy because the low energy particles must be produced with a large backward momentum in the center-of-mass reference frame, and so their progenitors are very high energy protons, whose intensity is relatively low.
This confirmed that, at least at low-energy, the secondary component is dominant in the CR $\bar{p}$ flux.

The $\bar{p}$ production spectrum is given by a convolution of the primary spectrum of nucleons with the differential $\bar{p}$ cross sections~\cite{2014JCAP...09..051K,2014PhRvD..90h5017D}, hence the predicted antiproton-to-proton ratio does not decrease with the same energy dependence as B/C. 

In fact, AMS-02 found that this ratio is almost flat at energies beyond $\sim 10$ GeV. 
However, it has been claimed that this behavior might still be consistent with pure secondary production once the relevant uncertainties affecting these computations are taken into account~\cite{2015JCAP...09..023G,2015JCAP...12..039E}.

\subsubsection{A possible (non-orthodox) interpretation of cosmic ray positrons and antiprotons spectra}
\label{sec:heterodoxy}

In most of the studies on Galactic cosmic rays, the main observations used to identify source and propagation effects has been the study of secondary nuclei (such as lithium, beryllium and boron) as outlined in Sec.~2 in this review.
The comparison of the spectra of different particle types, in particular $p$, $e^-$, $\overline{p}$ and $e^+$, that have different sources and different propagation properties,
can also be used to develop an understanding of the mechanisms that form the spectra.

\begin{figure}[bt]
\begin{center}
\includegraphics[width=10.0cm]{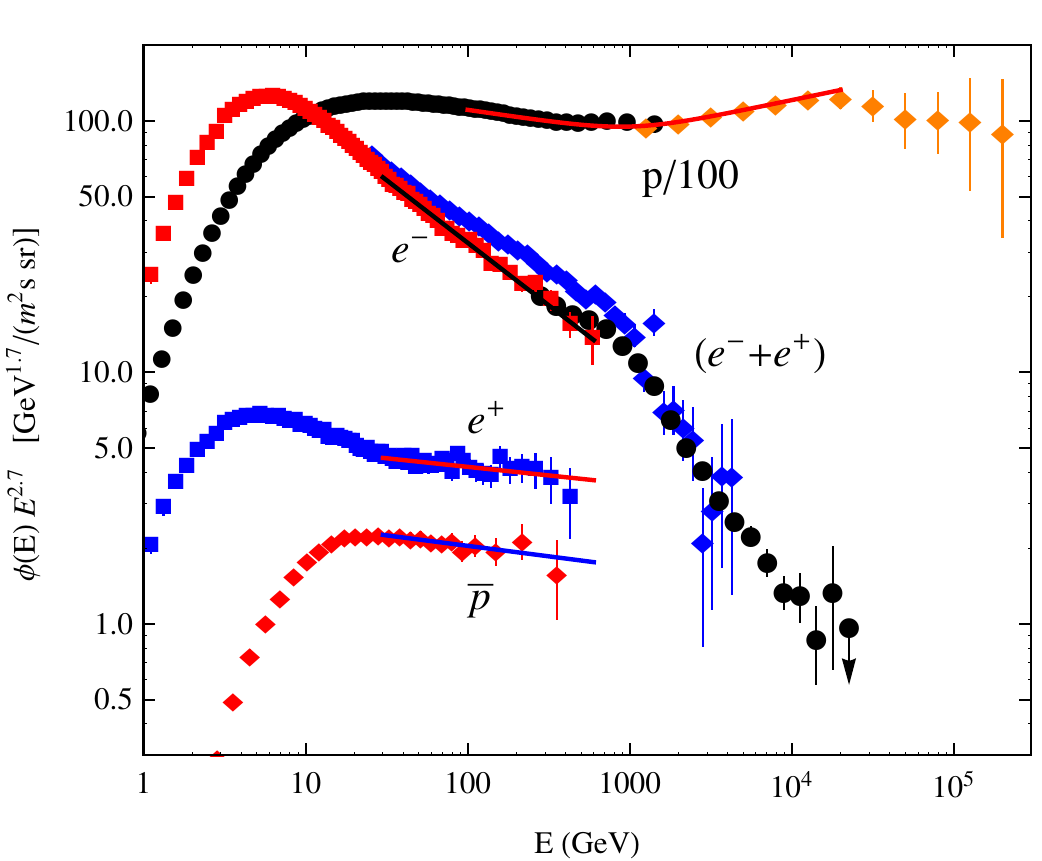}
\end{center}
\caption{Spectra of $p$, $e^-$, $e^+$, ($e^- + e^+$) and $\overline{p}$.
 The high energy data points for protons are from CREAM, the data points for
 ($e^- + e^+$) are from DAMPE and HESS. All other points are from AMS02.
 The lines superimposed to the $e^-$, $e^+$ and $\overline{p}$ data points
 are simple power law fits for $E > 30$~GeV.
 The line for the proton data is a broken power law fit. Figure from \cite{Lipari:2018usj}.
\label{fig:spectra_4particles}}
\end{figure}

Fig.~\ref{fig:spectra_4particles} show some recent measurements of the energy spectra for the four particles types which show several intriguing features that have to be understood.
One question that is obviously of central importance is the origin of the very different shapes (and normalizations) of the $p$ and $e^-$ spectra.
Intimately related to this question is the problem of the origin of the softening observed in the $(e^+ + e^-$) spectrum at $E \simeq 1$~TeV.

A second question that emerges naturally is why the spectra of $e^+$ and $\overline{p}$~(above an energy of approximately 30~GeV) have very similar shape, with spectral indices that are equal within errors.
This implies that the ratio of the two spectra is approximately constant with a value $e^+/\overline{p} \approx 2$ that, within systematic uncertainties, is equal to the expected ratio for the standard mechanism of secondary production.

The most commonly accepted interpretation for the difference between the $p$ and $e^-$ spectrum (what can be considered as the ``orthodoxy'') is that the difference emerges as the consequence of the effects of the much higher rate of energy loss for $e^\mp$ with respect to $p$ ($\overline{p}$), that softens an $e^-$ spectrum that at the source (when both particle types are ultrarelativistic) has the same shape as protons.

The idea that the difference in spectral shape between $e^-$ and $p$ is determined by propagation is simple and attractive, but there are some difficulties.
The first one is that since the $e^-$ spectrum is much softer than the $p$ spectrum already at an energy of few GeV, and the loss time for $e^\pm$ is of order $T_{\rm loss} \approx 620$~Myr/$E_{\rm GeV}$, this implies a very long residence time for CR in the Galaxy.

The second one is that the softening feature in the $e^-$ and $e^+$ spectra that should correspond to the transition between the regime where particle propagation is dominated by escape to the regime where energy loss is the dominant ``sink'' mechanism for the CR particles, that is now predicted in the GeV energy range, is not clearly visible,
perhaps because it is hidden by the spectral distortions created by solar modulation.
It should also be noted that the models have difficulties in describing the observed $e^-$ and $p$ energy distributions assuming source spectra of the same shape.

An additional important problem is associated to the interpretation of the prominent break in the ($e^- + e^+$) spectrum observed at 1~TeV. The origin of this structure remains controversial. An interesting possibility is that it is associated to the transition from escape to energy loss as the dominant effect in propagation. In this case however particles with a rigidity of order 1~TeV, should have a residence time of order 0.5--1~Myr, two orders of magnitude shorter than the previous estimate.

The properties of propagation in interstellar space for electrons and positrons (and for protons and antiprotons) are approximately equal.
This has a very important implications for the interpretation of the spectra of antiparticles ($e^+$ and $\overline{p}$).
If energy losses are the origin of the soft $e^-$ spectrum (and are therefore the dominant effect above an energy of few GeV), this implies that propagation effects should distort
the positron and antiproton source spectra (that have similar shape for $E \gtrsim 30$~GeV) in very different ways.

Why then do the positron and antiproton spectra have (in the energy range 30--500~GeV) spectra of very similar shape?
According to the ``orthodoxy'' this is simply a meaningless coincidence. The positron flux should be softened (relative to antiprotons) by propagation effects (approximately) as much as the electron flux is softened relative to the protons.
This then implies the existence of an additional hard source of positrons, that after the softening due to propagation results in a spectral shape equal to antiprotons. The absolute normalization of the new positron source is also such that the observed $e^+/\overline{p}$ is of order unity even if the origins of the two particle types are totally distinct.

The existence of these problems for the interpretation of the CR spectra, suggests the idea
to explore alternative ``heterodox'' scenarios based on the following ideas \cite{Lipari:2016vqk,Lipari:2018usj}: \\
(1) The difference in the observed spectra of $e^-$ and $p$ is generated in (or in the vicinity of) the sources, (presumably because of differences in the rate of energy loss), and not during propagation in interstellar space. \\
(2) The residence time of CR in the Galaxy is short (much shorter than in the ``orthodox'' explanation) of order 0.5--1~Myr at 1~TeV.
The propagation properties of $e^\mp$ and $p$($\overline{p}$) are therefore approximately equal for $E \lesssim 1$~TeV. \\
(3) The positron and antiproton fluxes are generated by the standard mechanism of secondary production, with no need for additional, non--standard sources. \\
(4) The break in the spectrum of the sum ($e^ -+e^+$) can be interpreted as the effect of energy losses during propagation. \\
(5) A comparison of the antiparticle spectra ($e^+$ and $\overline{p}$) with the spectra of protons and nuclei can be used to estimate the rigidity dependence of the CR Galactic residence time. This study results in a rather weak dependence, and therefore implies soft source spectra for protons and primary nuclei. \\ 
(6) The interpretation of the spectra of secondary nuclei (such as lithium, beryllium and boron) becomes now problematic and very likely requires production in (or in the vicinity of) the sources.

It should be stressed that there are significant difficulties in developing a consistent model that incorporate the ideas outlined above, but on the other hand there are also
significant difficulties in developing a consistent picture using the ``orthodox'' point of view.
The implications of the two different scenarios for cosmic ray astrophysics are profoundly different, because they result in very different properties for the sources and for the structure of the Galaxy, and therefore it seems important to study theoretically, and most importantly experimentally, these different scenarios.

Several observations have the potential to shed light on these problems: \\
(i) An extension of the measurement of the separate spectra of $e^-$ and $e^+$. \\
(ii) An extension of the energy range of the measurement of Beryllium isotopes. \\
(iii) The study of the space dependence of the spectra of electrons and positrons in the Galaxy. \\
(iv) A more accurate modelling of Solar Modulations. \\
(v) The identification of the primary CR sources, and the determination of their properties. \\
(vi) The identification of new sources of positrons (and antiprotons) if indeed they exist. 

The developments of programs of observations that can address the questions is obviously very desirable, and has the potential to lead to the clarification of these very important problems. 

It should be stressed that models where the imprints of energy losses on the electron and positron fluxes predict softenings of the two spectra at approximately the same energy.

Recently the AMS02 Collaboration \cite{Aguilar:2019owu} has released more data on the positron flux that extend the measurements to a maximum energy of 1~TeV.

The high energy points show that the $e^+$ spectrum has a marked softening that is fitted as an exponential suppression $e^{-E/E_s}$ with $E_s = 810^{+310}_{-180}$~GeV.

A possible interpretation this result is that the spectral suppression measures the maximum energy of a new hard source of positrons. On the other hand
(see \cite{Lipari:2019abu} for a more extended discussion) the spectrum could also be consistent with a ``break'' structure generated by energy loss effects. A comparison of the
softening structures observed in the $e^+$ and ($e^+ + e^-)$ spectra is essential to discriminate between these possibilities.

\subsection{Deficit in large-scale anisotropies and unexpected small-scale anisotropies}
\label{sec:anisotropies}

As discussed in Sec.~\ref{sec:orthodoxy}, the diffusive paradigm for the transport of Galactic CRs is based on two fundamental observations. First, the column depth of $\approx 10 ~\text{g}/\text{cm}^{2}$ inferred from secondary-to-primary ratios like Boron-to-Carbon requires CRs to traverse the gaseous disk a large number of times (see Sec.~\ref{sec:pillar2}). Second, the arrival directions of CRs are remarkably isotropic, down to one part in 10,000 at tens of TeV which implies the need for efficient randomisation of CR directions. If CRs were travelling ballistically, the anisotropy would be roughly $\sim 1 / \sqrt{N_{\text{src}}}$; even with $N_{\text{src}} = 10^6$ sources contributing this would fall short of the needed suppression~\cite{2005ppfa.book.....K}. Resonant interactions with a turbulent magnetic field, however, can be the agent of randomisation as it leads to efficient pitch-angle scattering. (The pitch angle is the angle between the CR momentum and the regular magnetic field.) In the limit of only small perturbations on top of a regular background magnetic field, the rate of scattering $\nu$ can be computed in a perturbative approach called quasi-linear theory~\cite{1966ApJ...146..480J,1966PhFl....9.2377K,1967PhFl...10.2620H,1970ApJ...162.1049H}. The rate of pitch-angle scattering is larger than the (relativistic) gyro frequency $\Omega_{\gamma}$ by the ratio of energy densities of the turbulent and the regular magnetic field, $\nu \sim (\delta B^2 / B_0^2) \, \Omega_{\gamma}$.

In quasi-linear theory, due to the low level of turbulence necessary for the validity of the perturbative approach, diffusion is markedly anisotropic. The pitch-angle diffusion induces a relatively large parallel diffusion coefficient, $D_{\parallel} \sim c^2 / (3 \nu)$ while the perpendicular diffusion coefficient, for instance in isotropic turbulence, is smaller by $(\delta B^2 / B_0^2)^2$, with $c$ denoting the speed of light. For moderate levels of turbulence other processes like field-line random walk will likely dominate the contribution to perpendicular diffusion. Formally, isotropic diffusion (with a diffusion tensor proportional to the identity matrix) is outside the validity range of quasi-linear theory, but oftentimes the normalisation and rigidity dependence of the diffusion coefficient are motivated from quasi-linear theory. Note that despite Galactic diffusion being \emph{locally} anisotropic, it can be \emph{effectively} isotropic on Galactic scales as the direction of the regular background magnetic field varies on Galactic scales~\cite{Ginzburg:1990sk}.

Despite the efficient randomisation, a residual anisotropy is retained if the distribution of CR sources is inhomogeneous. In that case, there will be a gradient in the isotropic part of the density, thus inducing a dipole in the arrival directions of CRs. In isotropic diffusion, the dipole points up the CR gradient, i.e.~towards the source(s). This has led to the suggestion to use CR anisotropies to search for nearby source(s)~\cite{1995ICRC....3...56P,Buesching:2008hr,DiBernardo:2010is,Borriello:2010qh,Linden:2013mqa}. In isotropic diffusion, the dipole amplitude $a$ is given by
\begin{equation}
a = \frac{3 D}{c} \frac{| \vec{\nabla} n_{\text{CR}} |}{n_{\text{CR}}} \, ,
\end{equation}
where $n_{\text{CR}}$ denotes the CR density. In the limit of anisotropic diffusion, the observed dipole will be aligned with the (local) field direction and the amplitude will be reduced by the cosine of the angle between gradient and background field, that is $\nabla n_{\text{CR}}$ is replaced by $\vec{\hat{b}} \cdot \vec{\nabla} n_{\text{CR}}$ with $\vec{\hat{b}} = \vec{B}_0 / | \vec{B}_0 |$. This projection limits the amount of information on sources encoded in the anisotropies. Another expected source of a dipole anisotropy is the Compton-Getting effect~\cite{CG1935,1968Ap&SS...2..431G} which is due to the relative motion between the observer and the frame in which the cosmic ray density is isotropic.

Whereas in the simplest diffusion models with a scattering rate constant in pitch-angle, only the dipole is induced, more realistic turbulence models can induce also higher multipoles~\cite{Giacinti:2016tld}. Furthermore, the stochastic nature of the turbulent magnetic field leads to some variance of dipole amplitudes and directions in different realisations of the turbulent magnetic field~\cite{Mertsch:2014cua}.

\begin{figure}[t]
\centering
\includegraphics[width=0.9\textwidth]{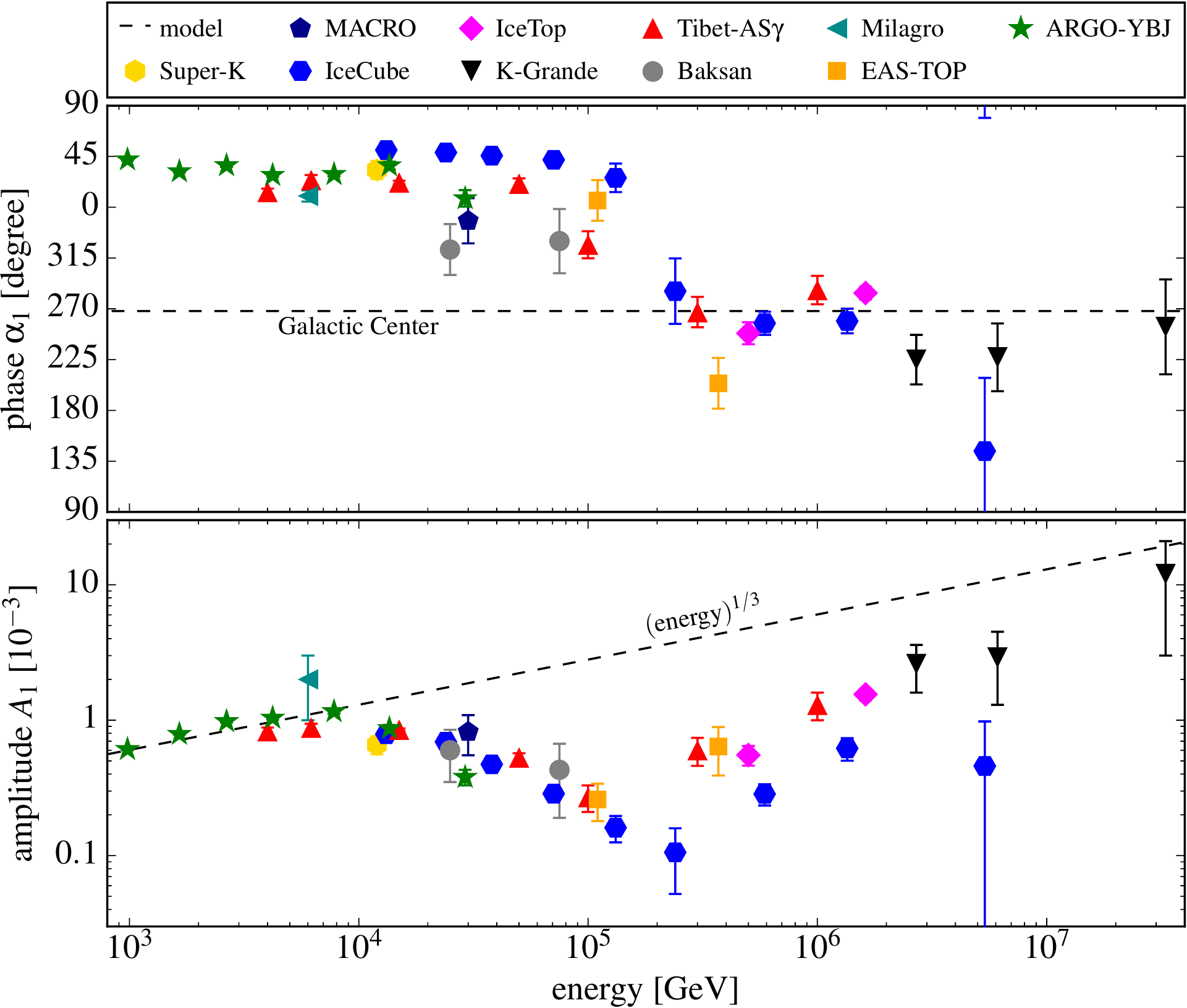}
\caption{Phase and amplitude of the (equatorial) dipole anisotropy from recent measurements. From Ref.~\cite{Ahlers:2016rox}.}
\label{fig:dipole_anisotropy}
\end{figure}

In Fig.~\ref{fig:dipole_anisotropy}, we show the phase and amplitude of the dipole anisotropy as recently measured by different experiments. Up to $\sim 10 \, \text{TeV}$, the amplitude is growing to $\sim 10^{-3}$, before decreasing and reaching a minimum of around $10^{-4}$ at a few hundred TeV. At even higher energies, it increases fast and reaches $\sim 10^{-1}$ at a few tens of PeV. The phase is relatively constant at $\sim 45^{\circ}$ right ascension before rapidly changing to $\sim 270^{\circ}$, the direction of the Galactic centre, at a few tens of PeV. The energy of this transition is very close to the energy of the amplitude minimum.

It has been noted a while ago~\cite{1973ICRC....1..500C} that the level of anisotropy is at variance with what is expected in simple diffusion models. This problem has become known as the CR anisotropy problem~\cite{2005JPhG...31R..95H}. Data collected over the last ten to 15 years have consistently confirmed the discrepancy.

The ``simple diffusion models'' make a number of assumptions, including: (1) Diffusion is isotropic and the diffusion coefficient is largely homogeneous. (2) The rigidity-dependence of the diffusion coefficient is as expected for a Kolmogorov or Kraichnan phenomenology of the turbulent magnetic field. (3) The distribution of sources is smooth and follows the Galactic distribution of supernova remnants or pulsars.

Various interpretations of and solutions to the anisotropy problem have been suggested. The fact that in anisotropic diffusion only the projection of the gradient onto the regular field direction matters can be used to reduce the tension by allowing for a misalignment between the gradient and the regular background field~\cite{Kumar:2014dma,Mertsch:2014cua}. In addition, the stochasticity of the turbulent magnetic field can produce realisations of the turbulent field that lead to lower dipole amplitudes~\cite{Mertsch:2014cua}. (In isotropic diffusion, this also weakens the directional association between gradient and dipole directions.) Furthermore, due to the discrete nature of CR sources in space and time, the density of CRs in the Galaxy is very sensitive to the exact distribution of young and/or nearby sources~\cite{Erlykin:2006ri,Ptuskin2006}. This results in variations as a function of position and time of the observed dipole amplitudes and phases. Modelling of the nearby known sources~\cite{Sveshnikova:2013ui,Ahlers:2016njd} and Monte Carlo simulations~\cite{Blasi:2011fm,Pohl:2012xs} show variations of the amplitude and phase. However, in order to suppress the amplitude to the observed level of $10^{-4}$ at tens of TeV would require for the dipole from the Galactic center to cancel the dipole from a nearby source in the opposite direction~\cite{Ahlers:2016njd}. One of the earliest solution to the anisotropy problem suggested~\cite{Cowsik:1975tj}, was to limit the rigidity-dependence of the diffusion coefficient (see also~\cite{Blasi:2011fm}). Finally, simple diffusion models can be overestimating the size of the CR gradient and thus also the dipole anisotropy~\cite{Evoli:2012ha}: If diffusive escape was increased in the inner Galaxy (via higher levels of self-induced turbulence and thus larger perpendicular diffusion) this would also help with the gamma-ray gradient problem, see Sec.~\ref{sec:gradient}.

The high statistics of observations performed over the last decade have not only improved the determination of the large scale anisotropies, but also led to a serendipitous discovery at TeV-PeV energies: A number of localised hot-spots at the level of $10^{-4}$ were found by Tibet~\cite{Amenomori:2006bx}, Milagro~\cite{Abdo:2008kr}, and ARGO-YBJ~\cite{ARGO-YBJ:2013gya}. IceCube extended the analysis of small-scale anisotropies to the southern hemisphere and presented their first angular power spectrum~\cite{Abbasi:2011ai}, proving that there is structure even on the smallest scales (as accessible with the given statistics). These pioneering observations have been confirmed and extended by HAWC and IceCube~\cite{Abeysekara:2014sna,Aartsen:2016ivj,Abeysekara:2018qho,TheHAWC:2017uyf}, showing evidence for a smooth angular power spectrum with anisotropies present down to angular scales of $\lesssim 10^{\circ}$.

\begin{figure}[t]
\centering
\includegraphics[width=0.8\textwidth]{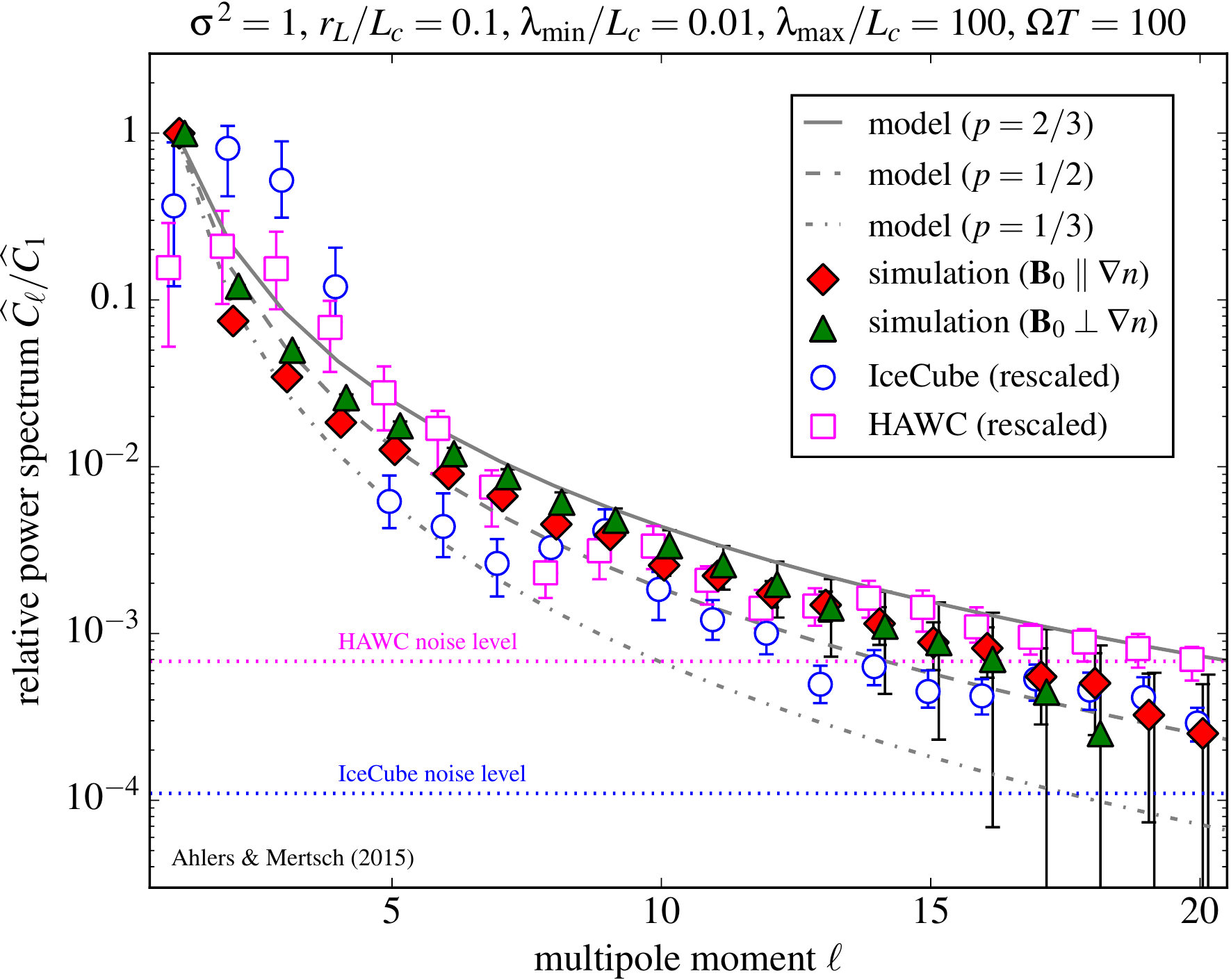}
\caption{The angular power spectrum of CR anisotropies. The open circles and squares show the measurements by IceCube~\cite{Aartsen:2013lla} and HAWC~\cite{Abeysekara:2014sna}. The lines are the result of an analytical calculation with the parameter $p$ determining the rate of decorrelation of trajectories. The diamonds and triangles show the numerical computation of the angular power spectrum and differ in the orientation of the CR gradient $\vec{\nabla} n$ and the regular background field $\vec{B}_0$. See Ref.~\cite{Ahlers:2016rox} for details.}
\label{fig:anisotropy_power}
\end{figure}

The appearance of small-scale anisotropies is most puzzling in the simple diffusion models that usually only predict dipole anisotropies (see, however, Ref.~\cite{Giacinti:2016tld}) and has led to a number of explanation. Historically the first were suggestions of unusual large-scale structures in the nearby Galactic magnetic fields, like magnetic lenses~\cite{Salvati:2008dx,Battaner:2010bd,Harding:2015pna}, that can lead to very narrow pitch-angle distributions. However, it was quickly pointed out~\cite{Drury:2008ns} that scattering would wash out such pitch-angle distributions and hence such structures needed to be closer than a mean-free path for which there is no observational support. A more elaborate idea is non-uniform pitch-angle scattering that can also generate a sharp feature in the direction of the regular background field~\cite{Malkov:2010yq}. However, the presence of hot and cold spots in various directions implies that this is likely not the full explanation. The fact that one of the originally observed hot spots is very close to the heliotail direction has led to speculation over the role of the heliosphere in shaping the anisotropy~\cite{Lazarian:2010sq,Desiati:2011xg,Drury:2013uka}. While some of the earlier models show some resemblance with the unusual large-scale structure models (see above), it has been shown in more sophisticated simulations~\cite{Schwadron2014,Zhang:2014dsu,Lopez-Barquero:2016wnt} that the heliosphere does play a role in shaping the anisotropies even if it is not the sole cause.

The arguably most elegant and economical explanation is to consider the role of the local, turbulent magnetic field in shaping the arrival direction of CRs~\cite{Giacinti:2011mz}. Whereas the standard diffusion models only consider the ensemble averaged phase-space density, small-scale anisotropies point to correlations between pairs of trajectories of locally observed CRs. Those correlations must be present to a certain degree as particles arriving under small angles will have experienced similar turbulent magnetic field orientations. The success of this idea was proven in a number of numerical~\cite{Giacinti:2011mz,Ahlers:2015dwa,Pohl:2015fdp,Lopez-Barquero:2015qpa} and analytical studies~\cite{Ahlers:2013ima,Ahlers:2015dwa}. See Fig.~\ref{fig:anisotropy_power} for a comparison between analytical and numerical results with data. Note that there are also a number of more exotic possibilities, like strangelets~\cite{Kotera:2013mpa} and dark matter substructure~\cite{Harding:2013qra}.

\subsection{Challenges from observations of far away cosmic rays}
\label{sec:indirect}

Observation of the diffuse gamma-ray emission from the Galactic disk can be used to obtain stringent constraints on the propagation of CRs in the Galaxy, as well as on the distribution of CR sources. Over the past decade, data obtained by Fermi-LAT provided us with an unprecedented view of our Galaxy in GeV gamma rays. The emerging picture contains unexpected features, which are described in the remainder of this Section.

\subsubsection{GeV excess}
\label{sec:gevexcess}

Over the latest decade, an increasing number of studies \cite{2009arXiv0912.3828V,Hooper:2010mq,Daylan:2014rsa,Macias:2013vya,Calore:2014xka,Abazajian:2014fta,TheFermi-LAT:2017vmf} focused on the inner Galaxy have repeatedly uncovered a statistically significant component in Fermi-LAT data, whose spectral and morphological features were initially claimed to be compatible with a dark matter pair annihilation signal. This component usually emerges out of template-fitting analyses, in which the gamma-ray sky is decomposed into different components of given morphology (most importantly point and extended sources, inverse Compton emission, hadronic emission, extra-Galactic background), whose normalizations are fitted to the data in each energy bin. This ``GeV excess'' is characterized by a bump-like spectral feature in the spectral energy distribution which peaks around $\simeq3$ GeV and shows an approximately spherical morphology. This discovery triggered a stimulating debate in the gamma-ray community and several alternatives, i.e.\ astrophysical interpretations, were put forward. Ref. \cite{Gaggero:2015nsa} and \cite{Carlson:2015ona} re-analyzed this anomaly by implementing a steady-state CR source term in the GC region that reflects the very high star-forming activity in the massive molecular complex known as {\it central molecular zone}: The inclusion of this term alters the IC template and absorbs most of the excess, significantly lowering the normalization of the GeV peak. The modeled IC template is also altered by the magnetic field model used for the propagation models. For example, a very recently work \cite{2019PhRvD..99d3007O} obtained IC templates more peaked towards the inner Galaxy region at high energies, by using propagation models that account for constraints of the Galactic magnetic field from non-thermal interstellar emission in radio and microwaves. Also importantly, several independent studies \cite{Bartels:2015aea,Lee:2015fea} implementing different statistical techniques designed to identify the granularity in the photon count maps have recently showed evidence for a population of unresolved point sources, possibly millisecond pulsars in the Galactic bulge. The recently developed \texttt{SkyFACT} \cite{Storm:2017arh} tool, based on both template fitting and image reconstruction techniques, has allowed for further improvement in the quality of the fits by implementing pixel-by-pixel variations of the Galactic diffuse emission templates: A recent investigation of the morphology of the GeV excess based on these techniques finds a correlation with the stellar distribution in the central boxy/peanut-shaped bulge/bar that characterize the inner regions of the Galaxy, and seems to provide more support to the millisecond pulsars hypothesis \cite{Bartels:2017vsx}. The excess has also been comprehensively analyzed by the Fermi-LAT collaboration \cite{TheFermi-LAT:2017vmf}. The bottom lines of this decade-long debate, in our opinion, are the following: (1) The existence of an extended signal from the inner Galaxy peaking at few GeV is fairly well established; (2) The precise characterization of this anomaly seems to depend quite significantly on the assumptions regarding the spatial distribution of CR sources; (3) The interpretation in terms of unresolved point sources will be testable in the future by means of more sensitive radio facilities \cite{Calore:2015bsx}.

\subsubsection{Progressive hardening in the hadronic gamma-ray emission}
\label{sec:hardening}

\begin{figure}[t]
\centering
\includegraphics[width=0.9\textwidth]{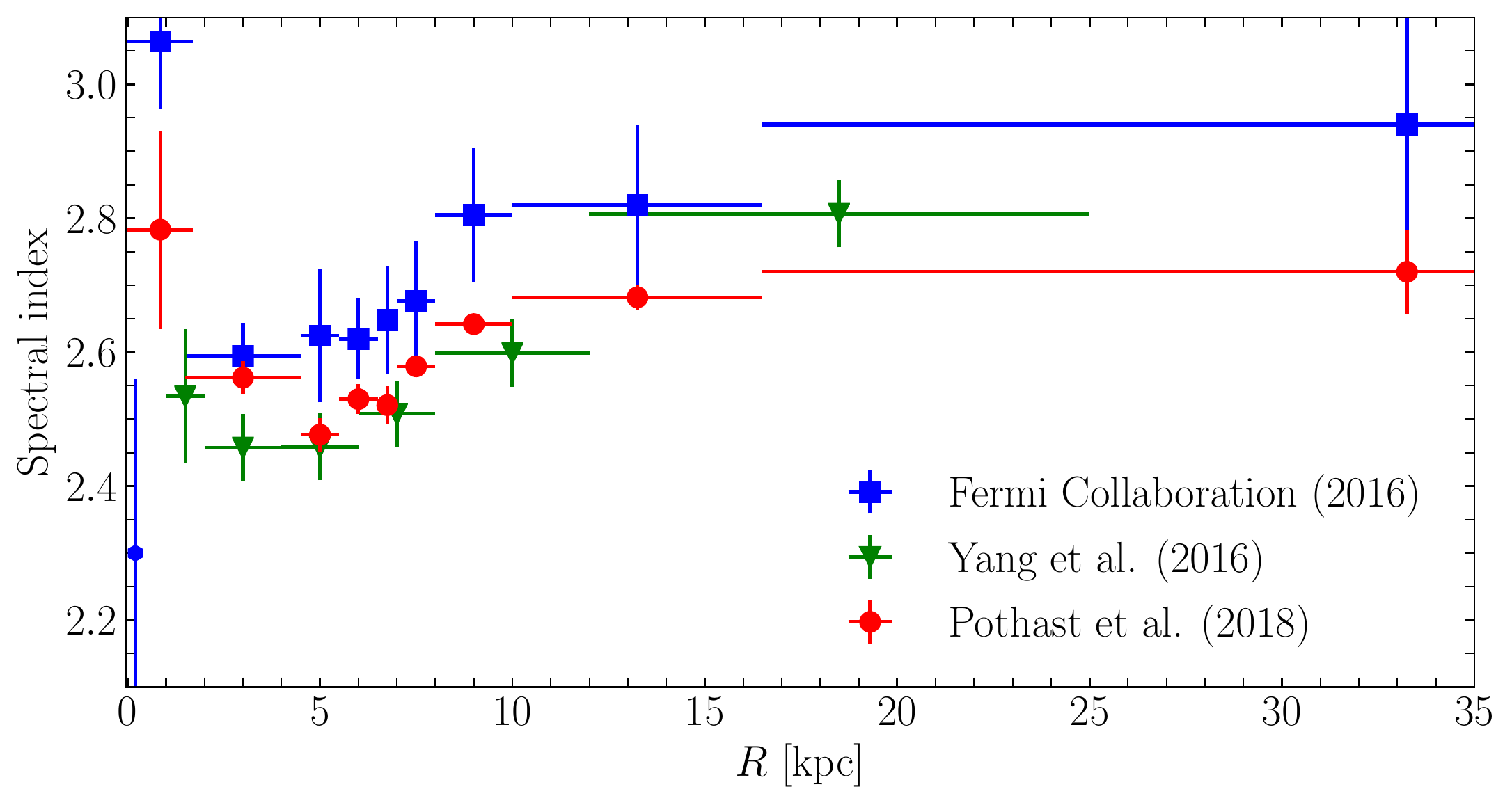}
\caption{The hadronic CR spectral index inferred by Fermi-LAT data, according to three recent analyses based on template-fitting techniques, see \cite{Yang:2016jda,Acero:2016qlg,Pothast:2018bvh}. The first point of the dataset taken from \cite{Acero:2016qlg} corresponds to the Central Molecular Zone only, which was cut out from the innermost ring and associated to a dedicated gas column density map. Figure adapted from \cite{Pothast:2018bvh}.}
\label{fig:hardening}
\end{figure}

The Fermi-LAT gamma-ray data offer the unique opportunity to probe the properties of the diffuse and ubiquitous flux of cosmic particles in different regions of the Galaxy and learn useful insights on their propagation in a quite broad range of energies.
This is a very challenging task since it requires, once the point and extended sources are properly subtracted, to disentangle the different types of emission.
A pioneering attempt to study the variation of the spectral index associated to the hadronic CR sea was presented in \cite{Gaggero:2014xla}: A progressive hardening of the CR spectrum towards the GC is outlined in that paper, and a phenomenological interpretation in terms of a spatially varying scaling of the diffusion coefficient with rigidity is presented. The presence of the trend was then better characterized by means of different template-fitting techniques in \cite{Yang:2016jda,Acero:2016qlg,Pothast:2018bvh}, as illustrated in Fig.~\ref{fig:hardening}. An even more recent analysis based on the study of individual molecular clouds is presented in \cite{2018arXiv181112118A}, where the existence of such trend is actually questioned.

The debate on the possible physical interpretations of this trend has then taken two different avenues.

On the one hand, Ref. \cite{Recchia:2016bnd} invokes a preeminent role of advection with respect to diffusion in the inner Galaxy, in particular in the region of the Molecular Ring, to explain the trend. This effect may be due to a more significant contribution of streaming instability in that region: Indeed, the growth of Alfv\'enic perturbations induced by the streaming of CRs themselves -- mostly taking place in the hot ionized phase of the interstellar medium in presence of strong gradients in the CR flux -- is expected to be stronger in that region due to the abundance of potential CR sources. 

On the other hand, Ref. \cite{Cerri:2017joy} devises an interpretation based on two pieces of information: {\bf 1)} The presence of a poloidal coherent component of the regular Galactic magnetic field directed in the perpendicular direction with respect to the Galactic plane in the inner Galaxy; {\bf 2)} The harder slope of the parallel (w.r.t. the regular field) diffusion coefficient with energy, hinted to by a low-energy extrapolation of numerical simulations \cite{DeMarco:2007eh,Snodin:2015fza}. Given these considerations, the authors developed a CR transport model based on anisotropic CR transport. The hardening, according to this interpretation, is caused by the progressive transition from perpendicular to parallel vertical escape from the Galactic plane: If confirmed, such a scenario would imply that the hard CR spectrum in the inner Galaxy extends all the way up to the multi-TeV domain, with relevant consequences on the interpretation of TeV gamma-ray data and on neutrino astronomy, see e.g. \cite{2008APh....30..180G,Gaggero:2015xza,Gaggero:2017jts}.

\subsubsection{Gradient problem}
\label{sec:gradient}

\begin{figure}[h]
\centering
\includegraphics[width=0.9\textwidth]{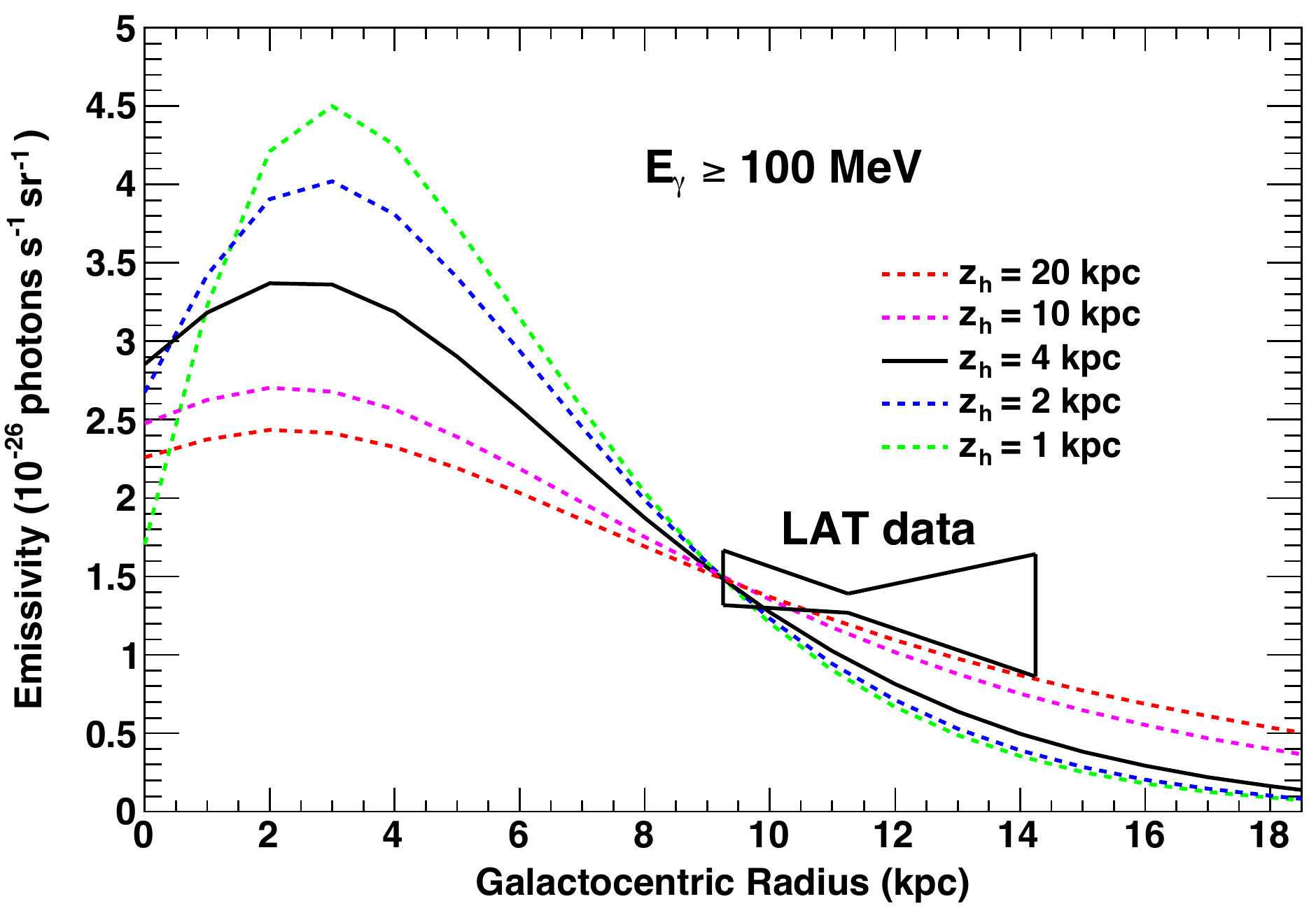}
\caption{The emissivity associated to the pion decay component of the diffuse gamma-ray emission as measured by Fermi-LAT and as predicted within standard CR propagation models. The different theoretical predictions correspond to different sizes of the diffusive halo of the Galaxy, as detailed in the legend. Only for very large values of this parameter the tension between model predictions and data is attenuated. Figure taken from \cite{Collaboration:2010cm}.}
\label{fig:gradient}
\end{figure}

How large is the gradient in the spatial distribution of CRs in the Galaxy?
Early observations by COS-B \cite{1988A&A...207....1S}, confirmed more recently and with increasing accuracy by EGRET \cite{2001ApJ...555...12D} and {\it Fermi}-LAT \cite{2010ApJ...710..133A,Collaboration:2010cm}, point towards a rather weak dependence of the CR proton density with respect to the Galactocentric radius in the outer Galaxy (i.e., for R $>$ 5 kpc). More precisely, the emissivity associated to the pion decay component of the gamma-ray emission appears to decrease by only 20\% to 40\% when moving from the Local to the outer spiral arm of the Galaxy. This poses a serious challenge to the standard theories of CR transport, where CR density is expected to trace the distribution of sources and therefore to rapidly decline beyond the solar circle. Such tension between observations and theoretical predictions goes typically under the name of \emph{gradient problem}.

The first attempts at a solution to the gradient problem did not involve any modification of the standard description of CR propagation. As an example, as investigated in \cite{Strong:1998pw}, one could reduce the steepness of the CR proton density by simply assuming a larger size for the diffusive halo of the Galaxy (10 kpc or more). This is shown in Fig.~\ref{fig:gradient}. However, matching the flatness of the density profile inferred from the gamma-ray observations appears to be rather difficult and, at some point, a very large halo size starts to be in tension with $^{10}$Be/Be and synchrotron data. An alternative solution would be to adopt a flatter distribution of sources, but this seems to be in contrast with SNR catalogs \cite{1998ApJ...504..761C}, pulsar catalogs \cite{2006MNRAS.372..777L}, and with the distribution of OB stars \cite{2000A&A...358..521B}. Another possibility, put forward in \cite{Strong:2004td}, lies in assuming a sharp rise of the factor $X_{CO}$ adopted to convert the intensity of the carbon monoxide line emission into the molecular hydrogen column density. Under this hypothesis, the flatness of the CR density in the outer Galaxy would just be the result of an erroneous determination of the molecular component of the interstellar hydrogen density in that region. However, the analysis performed by Fermi-LAT in \cite{Collaboration:2010cm} showed that the gradient problem is still present even if one considers only the diffuse emission resulting from collisions against the atomic component of the interstellar hydrogen.

A different approach in solving the gradient problem, which implies a modification of the standard description of CR diffusion, has been pursued in \cite{Evoli:2012ha}. Within this model, the diffusion coefficient perpendicular to the Galactic magnetic field is characterized by a spatial-dependent normalization, strongly correlated with the CR source distribution. One possible reason for this is that perpendicular diffusion is expected to be stronger in regions of high turbulence, where sources are more abundant. CRs thus escape more easily from these active regions and, as a result, the CR density profile is flattened. It is important to point out that such model could help with the large scale anisotropy problem, but not solve it as a smaller gradient doesn't necessarily lead to the very weak energy dependence between ~10 and 100 TeV (see Sec.~\ref{sec:anisotropies}).

\subsubsection{Fermi bubbles}
\label{sec:fermibubbles}

One of the most unexpected findings of the \textit{Fermi}-LAT mission was the discovery of the Fermi bubbles \cite{2010ApJ...724.1044S,2010ApJ...717..825D,2014ApJ...793...64A}. These huge, bi-lobular structures in gamma rays are likely emanating from the Galactic center and extend for $10 \, \text{kpc}$ above and below. They exhibit sharp borders and a constant surface brightness which is most peculiar for such a large and diffuse structure. Equivalently, their spectrum and intensity do not vary within the bubbles. Between $\sim 1 \, \text{GeV}$ and $\sim 100 \, \text{GeV}$, they have a $E^{-2}$ power law spectrum, thus harder than the Galactic diffuse emission. In principle, their gamma ray emission can be of hadronic or leptonic nature, but the likely association with the microwave haze~\cite{Dobler:2007wv,Ade:2012nxf} is more easily explained in leptonic scenarios. It is interesting to note that at high energy these structures are visible even without interstellar model subtraction. Also more recent synchrotron polarization maps \cite{2016A&A...594A..25P} show interesting features that may be correlated to the gamma-ray bubbles.

The directional association with the Galactic centre makes two sources for their energy very natural: a relativistic outflow, that is a jet, from past activity of the supermassive black hole (AGN scenario, e.g.~\cite{Guo:2011eg,Yang:2012fy}); or a period of star burst activity (or steady star formation) resulting in combined winds from explosions of massive stars (star burst/star formation scenario) (e.g.\cite{Crocker:2010dg,Sarkar:2015xta}). 

Broadly speaking, leptonic scenarios require fast transport of CRs if the acceleration sites are in or close to the disk due to the radiative losses on Galactic radiation fields and the CMB. An alternative is the \textit{in-situ} acceleration of electrons by stochastic acceleration in the bubble volume (e.g.~\cite{Mertsch:2018src}). Hadronic scenarios do not suffer such radiative losses and can thus maintain high-energy particles even if operating on much longer timescales, but particle confinement on Gyr timescales can become challenging. It is remarkable to see that CRs (hadrons and/or leptons) can be transported or accelerated up to a distance from the Galactic plane of $\approx 10$ kpc, which exceeds that commonly adopted for the Galactic halo ($\sim$ few kpc).

\subsubsection{CR spectra in the local interstellar medium}

When comparing gamma-ray diffuse emission observations with CR propagation models, models are usually based on direct CR measurements, which are assumed to be representative of CR properties throughout the local ISM. 

This assumption found support in the results of several early analyses aimed at inferring the local HI gamma-ray emissivity from Fermi-LAT data~\cite{2009PhRvL.103y1101A,2009ApJ...703.1249A,2012ApJ...750....3A}. These investigations focussed on intermediate latitudes (10$^o$~$<|$b$|<$~20$^o$), where the dominant contribution to gamma-ray emission is supposed to come from interstellar emission within $\sim$1 kpc from the Sun and where the uncertainties associated with CR propagation and gas distribution are expected to be minimized. Even with this considered, however, degeneracies in the gas distribution and interstellar radiation field, together with the contamination among the different components of the gamma-ray sky, can make the comparison between data and models challenging. Moreover, these first results were influenced by relatively large uncertainties both in gamma-ray data and in the directly measured CR spectra. 
 
A more recent study, illustrated in~\cite{2015ApJ...806..240C}, has led to a very accurate determination of the the local HI emissivity, using reprocessed Fermi-LAT data and more extended observations, based on the extensive analysis described in \cite{2016ApJS..223...26A}. 

This emissivity can be used to infer a local CR proton spectrum which can be compared with direct CR measurements, as discussed in \cite{2015ApJ...806..240C,2013arXiv1307.0497D,2015arXiv150705006S,2018MNRAS.475.2724O}. 
As shown in \cite{2018MNRAS.475.2724O}, the CR proton spectrum derived from the emissivity found in~\cite{2015ApJ...806..240C} is a factor of 1.4~$\pm$~0.5 larger than the very precise AMS-02 measurements at energies above 1 GeV \cite{PhysRevLett.114.171103}, even after including 10\% uncertainties in the cross sections. 

It is worthwhile to recall that earlier analyses, such as those described in \cite{2015ApJ...806..240C} and \cite{2015arXiv150705006S}, derived a similar proton spectrum, but no claim could be made due to the larger uncertainties of PAMELA data with respect to AMS-02 data used in \cite{2018MNRAS.475.2724O}.

The use of gamma-ray observations to investigate CR spectra in the local ISM is not limited to protons, but can be extended to electrons and positrons. This has been shown in \cite{2018MNRAS.475.2724O}, where, for the first time, the bremsstrahlung component of the local HI gamma-ray emissivity associated to CR electrons and positrons has been constrained by using observations of the synchrotron emission and CR direct measurements, by Voyager~1 \cite{Voyager} and AMS-02 \cite{PhysRevLett.113.121102,PhysRevLett.113.221102} in the low- and high-energy regime, respectively. The combined effect of the constraints coming from these different observables is that a local all-electron CR spectrum can be derived without the need of any assumption on solar modulation. Results obtained in this way show that both gamma-ray data and radio/microwave data prefer pure diffusion/convection models over the usual reacceleration models, as previously found in \cite{2011A&A...534A..54S} using radio/microwave data only, and confirmed in \cite{2013MNRAS.436.2127O} and \cite{2013JCAP...03..036D}.

In summary, present precise measurements of the local HI emissivity and very accurate CR direct measurement have started challenging also the assumption that CRs at Earth are representative of CRs in the local $\sim$1~kpc region. Moreover, both synchrotron (radio and microwave) and gamma-ray data challenge standard reacceleration models, in favor of diffusion models with no, or low, reacceleration.

\section{Long standing open problems in the standard paradigm}
\label{sec:problems}

The previous Sections have been devoted to a description of recent observational results which revealed unexpected features in the CR spectra, anisotropies, and spatial distribution. These findings questioned several of the aspects of the standard paradigm for the origin of CRs. Besides these new issues, several well known problems of the paradigm persists since many years. In the following, we list those which are, in out view, the most critical.
 
\subsection{The knee of the cosmic-ray spectrum}

The so-called "knee" is the most pronounced feature in the all-particle spectrum of hadronic CRs, and consists of a steepening of the spectrum to $E^{-3}$ at a particle energy of few PeV \cite{2005APh....24....1A}. For particle energies well below the knee the CR composition is largely dominated by protons, while for larger energies the presence of heavier elements becomes important. 
According to the usual consensus view, the spectral steepening observed above the knee corresponds to a reduced efficiency of Galactic CR sources in accelerating particles up to such energies \cite{2005A&A...429..755P,2006JPhCS..47..168H}. Alternative models are based on a reduced ability of the Galactic magnetic field to confine particles above the knee \cite{2015PhRvD..91h3009G}.

More specifically, in the usual consensus view the energy of the knee (few PeV) is associated to the maximum energy that protons can attain at Galactic accelerators. Heavier elements, being much less abundant than hydrogen in the ISM, are subdominant in the CR spectrum below the knee. However, since shocks accelerate particles according to their rigidity (and not energy), the contribution of heavy elements is expected to increase significantly {\it above} the knee, because the maximum energy that these elements can reach in a given accelerator is larger than that of hydrogen. Within this scenario, it has been shown that the sum of the contributions to the total CR spectrum from all the nuclear species reproduces well the $E^{-3}$ spectrum observed above the knee \cite{2006JPhCS..47..168H}.

Quantitatively, the Hillas criterion can be invoked to estimate the maximum energy of a particle of atomic number $Z$ at a SNR shock of size $R_{sh}$ moving at a velocity $u_{sh}$, giving \cite{2005JPhG...31R..95H}: 
\begin{equation}
E_{max} \sim \frac{Z}{3} \left( \frac{R_{sh}}{\rm pc} \right) \left( \frac{u_{sh}}{1000 ~\rm{km/s}} \right) \left( \frac{B_{up}}{\mu\rm{G}} \right) ~ \rm TeV
\end{equation}
where $B_{up}$ is the value of the magnetic field upstream of the shock. The rate at which SNRs convert energy into CRs peaks at the transition between the ejecta dominated and the Sedov phase, when the shock radius is of the order of the parsec or so, and the shock speed is several thousands of km/s. This implies that the predicted maximum energy of protons falls short of the knee \cite{1983A&A...125..249L} unless the magnetic field at the shock is dramatically amplified with respect to its interstellar value \cite{2004MNRAS.353..550B}.

Magnetic fields much stronger than interstellar ones are indeed observed in young SNRs \cite{2012A&ARv..20...49V}, and this could be a manifestation of a field amplification mechanism based on non-resonant CR streaming instability proposed by Bell \cite{2013MNRAS.431..415B}. However, even if this mechanism operates, it is not easy to amplify the field enough to have protons accelerated up to the knee at the time of the transition to the Sedov phase \cite{2008ApJ...678..939Z}. According to Bell \cite{2004MNRAS.353..550B}, the amplification is driven by the current generated by the CRs that, once accelerated up to $E_{max}$, escape the SNR. 
As a result, a fraction of the shock ram pressure $\varrho u_{sh}^2$ is converted into magnetic field energy density.
For a given CR acceleration efficiency at the shock (which must be about 10\% if SNRs are the sources of CRs), the effectiveness of the amplification mechanism will depend on the spectral slope of the accelerated particles (if the spectrum is steeper there will be less CRs at $E_{max}$ and therefore a weaker current and a less effective amplification) \cite{2013MNRAS.435.1174S,2014MNRAS.437.2802S,2015APh....69....1C,2016arXiv161007638G}. 
Reaching energies of the knee is problematic because, if one wants SNRs to accelerate CRs with a spectrum somewhat steeper than $E^{-2}$ (as required by observations, see Sec. \ref{sec:pillar3}), then the current of escaping CRs in not strong enough to provide enough amplification to allow protons to reach PeV energies.
On the other hand, if one wants SNRs to accelerate protons up to the knee, the required magnetic field amplification requires one to assume a spectrum of accelerated particles equal or harder than $E^{-2}$. 

From an observational point of view, conclusive evidence for the effective acceleration of PeV protons at SNR shocks would be provided by the the detection of neutrinos or gamma rays from such objects (or form their immediate proximity), with a spectrum extending unattenuated to the multi-TeV ($\gg 10$ TeV) domain \cite{2007ApJ...665L.131G}. 
However, the low statistics of spectra in the multi-TeV domain, together with the expected very small number of currently active SNR PeVatrons (at most a few in the entire Galaxy) make these observations challenging even for the most sensitive instruments \cite{2018MNRAS.479.3415C}.
Remarkably, the HESS Collaboration claimed the discovery of a "PeVatron" located in the vicinity (inner 10 pc) of the Galactic centre, and most likely linked to the central supermassive black hole \cite{2016Natur.531..476H}. This first detection is of great impact because it demonstrates that PeVatrons other than (the hypothetical) SNRs do exist in the Galaxy.

Thus, whether or not SNRs are capable of accelerating particles up to rigidities of few PV remains, to date, an open issue.

\subsection{The transition from Galactic to extragalactic Cosmic rays}

The {\it "ankle"} is a hardening in the CR spectrum observed at a particle energy equal to $\sim 3 \times 10^{18}$ eV. This hardening is believed to result from the emergence of the extragalactic component of CRs above the Galactic one \cite{2014NuPhS.256..197P}. As seen above, the knee in the CR spectrum is commonly considered to represent the maximum energy of protons accelerated at Galactic sources, with heavier nuclei shaping the CR spectrum at larger energies. Among heavy elements, the most abundant in the ISM is iron, characterized by an atomic number $Z_{Fe} = 26$. In the most standard scenarios for particle acceleration, nuclei are accelerated up to the same rigidity, and therefore the contribution to the total CR spectrum from Galactic sources should extend up to an energy of $E_{max}^{Fe} \sim Z_{Fe} E_{max}^H \approx 10^{17}$ eV, where $E_{max}^H$ is the energy of the knee. This energy is about 30 times smaller than the energy of the ankle, posing a problem to the standard scenario for the transition from Galactic to extragalactic CRs.

A possible solution would be to assume that the transition happens much earlier than the ankle, at an energy of $\approx 10^{17}$ eV. In fact, another (less pronounced) feature is observed at this energy in the CR spectrum. The feature is known as the {\it second knee} and appears as a slight steepening in the all particle spectrum \cite{2003APh....19..193H}. However, a transition at the second knee would require an unnatural amount of fine tuning, since the extragalactic contribution should appear sharply exactly where the Galactic one disappears (while in the case of a hardening in the spectrum such fine tuning is not required) \cite{2014NuPhS.256..197P}. Moreover, in this scenario one would also need to find another explanation for the ankle. Alternatively, one could assume the existence of a third CR population that would fill the gap between the SNR component and the extragalactic one \cite{2006JPhCS..47..168H}, or that a subset of the SNR population, rather than a new source class, fills the gap \cite{2010ApJ...718...31P}. For a critical analysis of these three-components scenarios, the reader is referred to \cite{2014NuPhS.256..197P}. 

Finally, another possible solution to the problem is to question the entire SNR paradigm and suppose that the contribution from sources other than SNRs dominates the CR spectrum all the way up to the ankle. According to the most popular alternative view, superbubbles rather than isolated SNRs may explain the entire spectrum of Galactic CRs up to the ankle \cite{2014NuPhS.256..197P,2014A&ARv..22...77B,2018arXiv180709726L,2018ARNPS..68..377T}. This scenario, however, surprisingly received much less attention than the SNR paradigm, and thus needs to be further investigated.

\subsubsection{Extragalactic CR: how far down in energy?}
\label{EGCR}

In a (definitely non-orthodox) paper in 1972 Brecher and Burbidge \cite{1972ApJ...174..253B} proposed that all (hadronic) CR\footnote{energy losses on the CMB exclude this for the leptonic component}
 are extragalactic (EG) and fill the universe at the same level as in the Galaxy.
The classical disproof of this conjecture was that the hadronic pion-decay gamma-ray background on intergalactic gas would exceed the observations.
Also the density of CR in the Galaxy seems vary with position \cite{2015PhRvD..91h3012G}, which would also argue against this idea.
However now 40 years later we have far better data and we ought to check up on the current situation.
The intergalactic gas density is about $10^{-7}$ H atoms cm$^{-3}$ from Big-Bang nucleosynthesis theory, and using measured Galactic gamma-ray emissivities \cite{2015ApJ...806..240C,2015arXiv150705006S} e.g. at 1 GeV: $2\ 10^{-24}$ MeV$^2$ sr$^{-1}$ s$^{-1}$ MeV$^{-1}$ H-atom$^{-1}$, we find (multiplying by the Hubble distance $\approx10^{28}$ cm) a diffuse background $2\ 10^{-3}$ MeV$^2$ cm$^{-2}$ sr$^{-1}$ s$^{-1}$ MeV$^{-1}$ compared to the observed intergalactic gamma-ray background IGRB (defined as the background excluding resolved extragalactic sources) $6\ 10^{-4}$ MeV$^2$ cm$^{-2}$ sr$^{-1}$ s$^{-1}$ MeV$^{-1}$ \cite{2015ApJ...799...86A}, so the resulting background is still too high but strangely close.
 Of course the IGRB is thought to be dominated by AGN and any residual diffuse flux much lower, so the excess is certainly robust and the universal CR theory is still excluded. 

Regarding the Galactic CR variations, the derivation of the Galactic CR density is fraught with uncertainty due to the difficulty of getting reliable densities of atomic and molecular hydrogen, including the so-called dark gas not traced by CO. So this is not a watertight argument against universal CR.
More solid is the relatively low interstellar flux from the LMC \cite{2016A&A...586A..71A} and M31 \cite{2017ApJ...836..208A}, less than expected if CR fill the universe.

Although there is evidence for hadronic production in SNR, this is not yet completely proved, as discussed in section \ref{PionBump}.

While it is unlikely that the universe is uniformly filled with CR, an extragalactic origin above say 1 TeV is certainly not excluded by any data. 
The IGRB has not been measured at these energies and there does not seem much prospect of doing so at present.

Most extragalactic CR models e.g.~\cite{2015PhRvD..92b1302G,2016A&A...595A..33T} invoke a quite hard CR spectrum e.g. $E^{-2}$ to avoid having a substantial EG component at lower energies, and EGCR are assumed to cut in above $10^{15}$ eV.
Instead a $E^{-2.5}$ law would give a few percent EGCR/GCR at 1 TeV and be consistent with available constraints.
Of course this would have to be investigated in the context of composition studies, secondary production etc.
But it does appear that restricting EGCR to above $10^{15}$ eV is a preconceived notion, again perhaps a truth universally accepted but not necessarily so.

The pair cascades induced by UHECR on the cosmological radiation fields are a current hot topic \cite{2016PhRvD..94d3008L,2016arXiv160903336V} because of their contribution to the extragalactic gamma-ray background, and it is interesting that apart from constraining models, there might be the prospect of eventually actually detecting this gamma-ray background among the other components, as first discussed by \cite{1973Natur.241..109S}.

\subsection{Chemical composition of cosmic rays} 
 
As seen in Section~\ref{sec:pillar3}, the chemical composition of CRs exhibits significant differences when compared to the solar one. Besides the large overabundance of light (LiBeB) and sub-iron elements in CRs, explained as the result of spallation, several less striking but still very significant differences exists between the composition of CRs and that of the (primordial) solar system. Most notably, CRs exhibit an overabundance of metals (nuclei with $Z > 2$) with respect of H and He, and of refractory elements over volatiles. While the former difference remains unexplained, the latter is most likely connected to the role played by dust grains during the acceleration process \cite{1997ApJ...487..197E}.

More recently, composition measurements by ACE at energies of hundreds of MeV/n~\cite{2014APS..APR.E9004B} and TIGER at GeV/n~\cite{2009ApJ...697.2083R} showed that the volatile abundance enhancement clearly depends on mass, while for the refractories this correlation appears to be less distinct. 
However, the ordering of these elements with atomic mass is greatly improved (with similar slopes) by comparing CR source abundances with a mixture of $\sim 80$\% of material with primordial solar composition and $\sim 20$\% of material enriched by stellar outflows and/or ejecta \cite{2016ApJ...831..148M}.

Such a mixed ISM can be found inside superbubbles which are inflated by the localized and episodic explosions of many supernovae in a stellar cluster \cite{2018arXiv180709726L}.
Moreover, the acceleration of CRs in metal enriched superbubbles, rather than in the typical ISM, might also provide an explanation for the abundances of Be and B (products of CR spallogenic nucleosynthesis) observed in old halo stars of different metallicity \cite{2000A&A...362..786P,2018ARNPS..68..377T}.
Unexpectedly, beryllium and boron abundances in metal poor halo stars are found to increase {\it linearly} with metallicity (expressed in terms of the abundance ratios [Fe/H] or [O/H]), rather than quadratically, as one would expect from a scenario where CRs are accelerated out of a standard (i.e. non enriched) ISM \cite{2000A&A...362..786P,2018ARNPS..68..377T}.

Some difficulties also persist in explaining the abundances of some CR isotopes, most notably the overabundance of roughly a factor of 5 of $^{22}$Ne with respect to expectations suggests that a significant fraction of CRs is accelerated out of Wolf-Rayet wind material, enriched in helium burning products \cite{1997ApJ...487..182M}. Whether this condition can be satisfied in superbubbles it is a matter of current debate \cite{2018arXiv180709726L,2012A&A...542A..67P}.
The same holds for the anomalous (enhanced) isotopic ratio $^{58}$Fe/$^{56}$Fe~\cite{2003ApJ...590..822H}.

Finally, a possible connection between CR acceleration and stellar clusters (OB associations) comes also from the recent detection of $^{60}$Fe in CRs. $^{60}$Fe is a $\beta$- unstable primary CR with half-life $\sim$2.6~Myr that is primarily synthesized in core-collapse SNe. Its presence in the cosmic radiation permits to set an upper limit of a few million years on the time between nucleosynthesis of these nuclei and detection at Earth, as it is the case if they were injected by a local (distance $\lesssim$~kpc) association of OB stars~\cite{2016Sci...352..677B}. Supported by the discovery of products of radioactive $^{60}$Fe in deep-sea sediments, this is evidence for a nearby and recent supernova, invalidating many of the current models of GCR.

\section{Conclusions}
\label{sec:conclusions}

Much of cosmic ray research in the past century has been devoted to answering a set of \textit{classical} questions:

\begin{enumerate}[I. ]
\item Which classes of sources contribute to the CR flux in different energy ranges? How many types of sources provide a significant contribution to the overall CR flux? 
\item Are CR nuclei and electrons accelerated by the same sources?
\item Which sources are capable of reaching the highest particle energies and how? 
\item Which are the relevant processes responsible for CR confinement in the Galaxy? 
\item Where is the transition between Galactic and extra-Galactic CRs and how can we explain the well-known features such as knee, second knee, ankle?
\item What is the origin of the difference between the chemical composition of CRs and the solar one?
\end{enumerate}

Many decades of direct and indirect observations of Galactic CRs, together with extensive theoretical investigations, have led to the development of a broadly accepted scenario for CR origin, which we have referred to before as the ``orthodoxy'' (see Sec.~\ref{sec:orthodoxy}).
Within this framework, plausible (though not conclusive) answers to most of the above questions can be obtained.

Thanks to impressive progress on the experimental side over the past $\approx$ 15-20 years, both in direct (Sections \ref{sec:spectral} and \ref{sec:anisotropies}) and indirect observations (Sections \ref{sec:test} and \ref{sec:indirect}), an enormous amount of data of unprecedented quality has allowed studying Galactic CRs in greater detail. Overall, a much more complicated picture than previously thought has emerged: As discussed in Section~\ref{sec:recentobservations}, new observations revealed a great number of anomalies and unexpected behaviors thus producing a whole list of new puzzles:

\begin{enumerate}[1. ]
\item What is the origin of the hardening observed in the spectra of CR nuclei at a rigidity of $\sim 300$ GV? (Section \ref{sec:nuclei})
\item Why is the slope of the spectrum of CR proton and helium different? (Section \ref{sec:nuclei})
\item What is the origin of the prominent break observed at a particle energy of $\sim 1$ TeV in the electron spectrum? (Section \ref{sec:electrons})
\item Why do the proton, positron, and antiproton spectra have roughly the same slope at particle energies larger than $\sim 10$ GeV? (Section \ref{sec:anti})
\item What is the origin of the rise in the positron fraction at particle energies above $\sim 10$ GeV? (Section \ref{sec:anti})
\item What is the origin of small scale anisotropies? (Section \ref{sec:anisotropies})
\item Why is the CR flux very close to isotropy up to very large particle energies? Why is the phase of the anisotropy pointing away from the Galactic centre for particle energies below $\sim 100$ TeV? (Section \ref{sec:anisotropies})
\item Can we explain the spatial variations of the CR intensity in the Galaxy? In other words: why is the spatial gradient of CRs so small? Why is the CR spectrum hardening towards the centre of the Galaxy? (Section \ref{sec:gradient} and \ref{sec:hardening})
\item What is the origin of the GeV excess detected from a region roughly coincident with the Galactic bulge? (Section \ref{sec:gevexcess})
\item What is the origin of the Fermi bubbles? (Section \ref{sec:fermibubbles})
\item Given the quite complicated picture emerging from data, how can we possibly use CRs to search for new physics? Is there a really ``clean'' channel? (For a more detailed discussion on this point, see for instance the recent review \cite{GaggeroValliRev}) 
\end{enumerate}

These new questions are much more detailed that the ``classical'' questions, reflecting of course the dramatic improvement in the quality of data that became available in recent times. Also, the amount of detail we can now observe challenges the currently accepted paradigm, which indeed did not predict any of the features found in the latest data. It is possibly for this reason that several alternative ideas to the SNR paradigm, proposed years ago to explain the origin of CRs, have been recently revived. The list includes the acceleration of particles at the Galactic centre (see \cite{1981SvA....25..547P} for an early reference, \cite{2016Natur.531..476H} for a recent one), at stellar winds (see \cite{1983SSRv...36..173C} for an early reference, \cite{2018arXiv180402331A} for a recent one), in superbubbles (see \cite{1992MNRAS.255..269B,1998ApJ...509L..33H} for early references, \cite{2014A&ARv..22...77B,2018arXiv180709726L,2018ARNPS..68..377T} for recent ones), or the explanation of at least part of the CR spectrum with a single (or very few) sources (see \cite{1997JPhG...23..979E} for an early reference, \cite{2019JCAP...01..046B} for a recent one).
Finally, other classes of sources (such as X-ray binaries \cite{2002A&A...390..751H,2005MNRAS.360.1085F} or pulsar wind nebulae \cite{2014IJMPS..2860160A}) have been considered, and attempts to develop truly heterodox scenarios were published, as we briefly discussed in Section \ref{sec:heterodoxy}.

In the light of these recent developments, it is mandatory to define an observational strategy for the future of CR research. For this reason, we list below 10 future direct and indirect observations of CRs that we consider crucial in order to solve the problem of their origin.

{\bf Direct observations of CRs}
\begin{enumerate}[1. ]
\item{\bf Extension of the measurements of $^{10}Be$ and $B/C$ to larger energies} -- Measurements of $^{10} Be$ and $B/C$ provide us with estimates of the total residence time of CRs in the Galaxy (disk plus halo), and of the grammage accumulated by CRs during such time. The simultaneous knowledge of both these quantities is required in order to constrain precisely the transport properties of CRs of a given energy. To date, measurements of $^{10}Be$ are limited to energies of $\lesssim 1$ GeV/nucleon, while the $B/C$ ratio is known up to $\lesssim 1$ TeV/nucleon. It is thus mandatory to measure $^{10}Be$ at larger energies. Also, the extension to the TeV domain of the measurement of $B/C$ will provide information on the transport properties of multi-TeV CRs (currently virtually unconstrained). Note that the grammage accumulated by CR inside or in the vicinity of their sources might be non-negligible in the very high energy domain. So, measurements of $B/C$ in the TeV domain might provide insights on the confinement and escape of CRs from their accelerators.
 
\item{{\bf Extension to larger energies of the measurements of the separate electron and positron spectra} -- The energy spectra of electrons and positrons for $E$ larger than 1 TeV carry very important information about their production mechanisms, the space distribution of the sources, (that could be different for $e^-$ and $e^+$) and the properties of CR propagation. The measurements of the magnetic spectrometer AMS extend to a maximum energy $E \approx 1$~TeV. At higher energy, valuable measurements of the flux for the sum ($e^- + e^+$) have been obtained by calorimeters (in space) and \v{C}erenkov telescopes, however, the extension of the measurements of the separate spectra in the multi--TeV energy range could be of great importance to clarify the mechanisms that form the $e^\mp$ spectra.} 

\item {\bf Extension to larger energies of the measurements of the antiprotons} -- This observable is currently measured with high accuracy by the AMS-02 collaboration. As mentioned in Section \ref{sec:anti}, AMS-02 initially claimed the presence of an excess above 100 GeV; immediately after, more refined analyses \cite{2015JCAP...09..023G,2015JCAP...12..039E} pointed out only a mild ($\simeq 2\sigma$) deviation with respect to the expected background from pp collisions in the interstellar gas, computed within the standard framework, well constrained by B/C and other secondary/primary ratios. 
Still, the hint deserves to be better characterized, and more accurate high-energy data are needed. If confirmed, this anomaly may point towards a non-negligible secondary production at the accelerators \cite{2009PhRvL.103e1104B,Mertsch:2014poa,cholis2017}. Spatial-dependent diffusion scenarios, for instance two-zone models as the ones described in \cite{feng2016prd,Guo:2018wyf} --- which seem to capture in a phenomenological way both CR transport within pre-existing Galactic turbulence and self-confinement due to streaming instability --- seem to solve the discrepancy as well. 
(For a discussion about possible dark matter implications, we refer to the recent review \cite{GaggeroValliRev})

\item {\bf Measure the anisotropy of both CR nuclei and electrons.} Future measurements of the anisotropy in the (dominant) nuclear component of CRs by existing (e.g.\ IceCube and HAWC) and future experiments (e.g.\ LHAASO) will greatly increase statistics, but also extend the energy range, thus enabling the study of the energy-dependence of small-scale anisotropies. It has been argued (without a detailed theoretical study though) that the interpretation in terms of magnetic turbulence should lead to rather fast variations of the anisotropy patterns with energy. Thus energy-dependence might be the smoking gun in the interpretation of small-scale anisotropies. In addition, observatories that can also target the (subdominant) component of CR electrons (and positrons) have the potential of setting bounds on (if not detecting) an anisotropy. While the electron channel might be much more strongly limited by statistics, the level of anisotropy could likely be larger, due to the stronger impact of a few, nearby sources. Both a detection or stronger bounds would allow important inferences on the origin of the positron excess.
\item {\bf Improved measurements of CR chemical composition, especially in the knee-ankle region} -- Studies of the chemical composition of CRs play a key role in the understanding of the origin of these particles, and should thus be encouraged at all particle energies. In particular, an improved measurement of the CR chemical composition in the knee and ankle regions would be particularly useful, in order to clarify which is the maximum particle energy achievable in Galactic accelerators, and to explore the transition from Galactic to extragalactic CRs. 
With this respect, ISS-CREAM and HERD from space are expected to explore the knee region for the first time with direct measurements, but also indirect experiments, as e.g. LHAASO, are promising since they are expected to provide a superior sensitivity.\\

\end{enumerate}

{\bf Indirect observations of CRs}
\begin{enumerate}[1. ]
\item{{\bf Improved gamma-ray observations of potential CR accelerators, especially in the multi-TeV domain}} -- Our knowledge on particle acceleration processes in SNRs and other astrophysical objects improved dramatically with the advent of the present generation of gamma-ray instruments (both ground based and space borne). Future instruments characterized by a superior sensitivity, and by improved angular and energy resolutions will allow, on one hand, to study in much more detail already known sources and, on the other, to increase dramatically the number of known sources. Having a better view of already known gamma-ray sources will translate into more stringent constraints on the acceleration processes operating at such objects. Detecting more sources will make possible population studies, aimed at identifying the global CR output from given classes of sources. Thit task will be certainly performed by CTA in the TeV domain, but hopefully also by other ground based and, in the GeV domain, by space born instruments. Observation in the multi-TeV domain, currently poorly explored due to lack of sensitivity, should be apriority, due to their importance in connection to the search for CR PeVatrons.
\item {\bf Measure the Galactic diffuse gamma-ray emission above TeV energies.} This very challenging task is currently pursued by the HAWC observatory \cite{HAWC2015IAUS..313...70L,HAWC2017ICRC...35..689Z}. In the future, the \v{C}erenkov Telescope Array (CTA) \cite{2017arXiv170907997C} -- currently under construction -- and forthcoming air shower arrays such as LHAASO \cite{2016NPPP..279..166D} and possibly a Southern Hemisphere extension to HAWC, will provide a very significant increase in sentitivity and angular resolution in the multi-TeV regime. 
CTA in particular has the opportunity to focus on the emission from large individual clouds in the Galactic ridge and across the whole Galactic plane. The analysis of multi-TeV diffuse emission will be of paramount importance to confirm or constrain the spectral trends identified in Fermi-LAT data detailed in section 3.3.2 and understand the mechanism of CR confinement in the TeV - PeV domain. 
\item {\bf Measuring the Galactic diffuse gamma-ray emission at MeV energies and below} -- Proposed missions at MeV energies, such as AMEGO\footnote{https://asd.gsfc.nasa.gov/amego/}, e-ASTROGAM \cite{eastrogam} (and the All-Sky ASTROGAM), COSI\footnote{http://cosi.ssl.berkeley.edu}, and GalCenEx\footnote{Moiseev's talk at the 12th INTEGRAL Conference and AHEAD Gamma-ray workshop, Feb 2019, Geneva} would allow to access a variety of features --- including in particular the 511 keV line from $e^+e^-$ annihilation, and the nuclear de-excitation lines --- and to probe the low-energy CRs, their distribution, density, and spectral information throughout the Galaxy. The latter aspect is particular important to the aim of understanding the CR interplay with the interstellar medium: for instance, the interactions of low-energy CRs (that are responsible for the heating and ionization of the interstellar gas), and the transport properties of the low-energy CR population filling highly complex regions such as superbubbles. For more details see for example the white paper on ISM and CRs with gamma-ray observations at MeV energies submitted to the Astro2020 Decadal Survey \cite{WhitePaperISM}. %(Feel free to add/change)
\item {\bf Neutrino observations in the southern sky} -- Neutrino observations will allow for an unambiguous identification of the sites where Galactic CRs are accelerated. In the upcoming future, the detection of neutrino point sources will be among the primary goals of the KM3NeT/ARCA detector \cite{Adrian-Martinez:2016fdl}, which will observe the southern sky (in the track channel) with unprecedented accuracy. A joint effort carried on by both current and future collaborations can also possibly identify a diffuse component associated to the Galactic CR population (see for instance the recent joint analysis by ANTARES and IceCube \cite{ANTARESIceCube2018ApJ}): As already mentioned in the context of TeV gamma rays, such a detection can shed light on scenarios characterized by different hadronic CR spectra in different regions of the Galaxy \cite{Gaggero:2015xza} (Section \ref{sec:gradient})
\item {\bf Solar modulation or Voyager-like observations} -- The shapes of the CR spectra at low energy ($E \lesssim 10$~GeV) also carry very valuable information about the CR sources and the properties of Galactic propagation. For example, according to many models, the signature of energy losses on the $e^\mp$ spectra should appear at particle energies of order few GeV. Also, in models where CR positrons are generated by a new non--standard source, the transition energy where the new source becomes dominant is of order 10~GeV. In this low energy range the effects of solar modulations are however important, and to extract information about the interstellar spectra from observations performed near the Earth it is necessary to have a sufficiently good understanding of the solar effects.
Measurements of the CR spectra over an extended period of time, that covers a broad range of solar activity states, and a comparison with the measurements obtained by Voyager at the boundary of heliosphere, are necessary to construct and test models for the solar modulation effects.
This is not only a valuable program in itself (with the heliosphere a smaller scale ``laboratory'' for the study of Galactic propagation), but would allow to interpret correctly the CR spectra also at low energy.
\end{enumerate}

In addition to all of the above, it should be also noted that the unprecedented quality of available CR data requires an equally accurate knowledge of the cross sections describing CR interactions. Their availability is mandatory in order to interpret correctly data, and an experimental work in this direction has to be considered as a priority \cite{2018JCAP...07..006E,2018PhRvC..98c4611G}.

We conclude by reminding that the goal of this review was to provide a critical analysis of the standard (orthodox) scenario for CR origin which is broadly accepted in our community. The key aspects of this scenario (shock acceleration at SNR shocks plus diffusive confinement of CRs in a magnetized halo) have been reviewed, and it has been shown how its predictions are questioned by a number of recent (and not so recent) observations.
Even though, to date, it is not possible to say whether a radical change of paradigm is needed, rather than a minor modification of orthodoxy, we believe that a critical attitude as well as the development of alternative or even heterodox scenarios will greatly benefit the quest for the origin of CRs.

\section*{Acknowledgments}

This research was supported by the Munich Institute for Astro- and Particle Physics (MIAPP) of the DFG cluster of excellence ``Origin and Structure of the Universe'' (The high energy Universe: gamma ray, neutrino, and cosmic ray astronomy, 26th February-23rd March 2018).
SG acknowledges support from Agence Nationale de la Recherche (grant ANR- 17-CE31-0014) and from the Observatory of Paris (Action F\'ed\'eratrice CTA).
CE acknowledges the European Commission for support under the H2020-MSCA-IF-2016 action, Grant No.~751311 GRAPES Galactic cosmic RAy Propagation: An Extensive Study. 
EO acknowledges support from NASA Grant No.~NNX16AF27G. 
DG has received financial support through the Postdoctoral Junior Leader Fellowship Programme from la Caixa Banking Foundation (grant n.~LCF/BQ/LI18/11630014).

\bibliographystyle{ws-ijmpd} 
\bibliography{manuscript}

\end{document}